\documentstyle[12pt,psfig]{article}

\newcommand{\resection}[1]{\setcounter{equation}{0}\section{#1}}
\newcommand{\appsection}{\addtocounter{section}{1} \setcounter{equation}{0}
             \section*{Appendix \Alph{section}}}

\def\EQ{\begin{equation}}
\def\EN{\end{equation}}
\def\bea{\begin{eqnarray}}
\def\eea{\end{eqnarray}}

\def\b {\beta}
\def\D{\Delta}

\def\bh{\bar{h}}

\def\t{\theta}

\def\bd{\begin{displaystyle}}
\def\ed{\end{displaystyle}}
\def\ba{\begin{array}}

\def\ea{\end{array}}
\def\EQ{\begin{equation}}
\def\EN{\end{equation}}
\def\bea{\begin{eqnarray}}
\def\eea{\end{eqnarray}}
\def\beano{\begin{eqnarray*}}
\def\eeano{\end{eqnarray*}}

\def\D{\Delta}

\def\D{\Delta}

\begin{document}
\oddsidemargin 5mm
\setcounter{page}{0}
\newpage     
\setcounter{page}{0}
\begin{titlepage}
\begin{flushright}
ISAS/EP/97/85
\end{flushright}
\vspace{0.5cm}
\begin{center}
{\large {\bf Reflection Scattering Matrix of the Ising Model \\
in a Random Boundary Magnetic Field}}\footnote{Work done under 
partial support of the EC TMR Programme {\em Integrability, 
non--perturbative effects and symmetry in Quantum Field Theories}, grant
FMRX-CT96-0012} \\
\vspace{1.5cm}
{\bf 
A. De Martino$^{a,b}$, M. Moriconi$^c$ and  
G. Mussardo$^{a,b,c}$} \\
\vspace{0.8cm}
$^a${\em International School for Advanced Studies, Via Beirut 2-4, 
34014 Trieste, Italy} \\ 
$^b${\em Istituto Nazionale di Fisica Nucleare, Sezione di Trieste}\\
$^c${\em International Centre for Theoretical Physics}\\
{\em Strada Costiera 11, 34014 Trieste,Italy}\\
\end{center}
\vspace{6mm}
\begin{abstract}
\noindent
The physical properties induced by a quenched surface magnetic field in the
Ising model are investigated by means of boundary quantum field theory in 
replica space. Exact boundary scattering amplitudes are proposed and used 
to study the averaged quenched correlation functions. 
\end{abstract}
\vspace{5mm}
\end{titlepage}
\newpage

\setcounter{footnote}{0}
\renewcommand{\thefootnote}{\arabic{footnote}}

\resection{Introduction}

There has recently been increasing interest in the exact estimations 
of boundary effects in statistical models and quantum field theories,  
following seminal works by Cardy based on conformal invariance 
\cite{Cardy} as well as a pioneering paper by Ghoshal and 
Zamolodchikov \cite{GZ}, based on the integrability of the boundary 
interaction\footnote{See also \cite{KF} for boundary scattering theory 
and refs.\,\cite{Binder,Diehl} for a general introduction 
to boundary effects.}. New results have been obtained in many subjects, 
as for instance, the subjects of quantum impurity problems and 
dissipative quantum mechanics \cite{Affleck,Saleur,LSS} or the exact 
calculation of correlation functions and Casimir energy in the presence 
of boundary conditions \cite{Cardy,KLM,Chat,TBAb}. Aim of this paper 
is to enlarge the range of applicability of boundary field theory 
to systems which present boundary effects induced by a quenched 
surface disorder and to propose an exact non--perturbative approach 
based on boundary scattering theory\footnote{We refer the reader to the
references \cite{GZ,ZZ,GM} for all details and notations relative 
to exact boundary and bulk $S$--matrices.}. The model considered here is
the two--dimensional Ising model in the euclidean half--plane 
$\Gamma=\{(x,y)\in {\bf R};x \geq 0\}$ with a random magnetic field coupled
only to the boundary spins. This model was previously investigated by 
renormalization group techniques and conformal field theory methods in
\cite{Cardyb,Turban}. Its dynamics may be described by the euclidean 
action
\EQ
{\cal A} = {\cal A}_0 + \int_{-\infty}^{+\infty} dy \,h(y) \sigma(y) \,\, ,
\label{action}
\EN
where ${\cal A}_0$ is the bulk action relative to the Ising model in its
high--temperature phase in a half--space geometry, with free boundary
conditions adopted at $x=0$. In eq.\,(\ref{action}) $\sigma(y)$ is the local
boundary magnetic operator and $h(y)$ is a quenched random field with 
Gaussian distribution of mean $\bar h$ and variance $\Delta$:
\EQ
\overline{h(y)} = \overline{h} \,\,\,\,\, ; 
\hspace{5mm} 
\overline{h(y_1) h(y_2)} = \overline{h}^2 + \Delta \, \delta(y_1-y_2) \,\,\,.
\label{moments}
\EN
In the bulk, the massive excitation of the model may be regarded as an
interacting bosonic particle $A(\theta)$ created by the magnetization
operator\footnote{As usual, $\theta$ is the rapidity variable which 
parameterises the dispersion relations $(E,p) = 
(m \cosh\theta,m \sinh\theta)$ of the particle.}. Its elastic $S$-matrix 
is simply given by $S = -1$ \cite{KS}. The case $\Delta =0$ corresponds 
to the usual Ising model with boundary magnetic field, which was originally 
discussed by Ghoshal and Zamolodchikov \cite{GZ}: these authors proved that
the model is integrable and therefore all its properties (as for instance, its 
partition functions and its correlators \cite{KLM,Chat}) can be 
recovered by using both its exact 2--body $S$-matrix in the bulk 
and the elastic boundary scattering matrix given by  
\EQ
{\cal R}(\theta,\bh) = i \tanh\left(\frac{\theta}{2} - 
\frac{i \pi}{4}\right)  
\frac{i \sinh\theta  - \kappa}{i
\sinh\theta + \kappa} \,\, ,
\label{gz}
\EN 
where $\kappa = 1-\frac{{\bh}^2}{2 m}$. 

In terms of microscopic processes, the scattering event is nothing else 
but the result of an infinite number of interactions which take place on 
the boundary and which enter the self--energy of particle $A$, each of
these interactions involving the insertion of the magnetic field $h$ 
(see for instance Fig.\,1.a for a graph of order $h^6$). The integrability
of the model means that all scattering processes at the boundary proceed
without particle production or absorption, i.e. an observer will always 
see a particle arriving at the boundary and bouncing back. 

The graph technique can be also used to account for the interaction 
due to the random fields. The only new feature is that we have 
to average over the random variables $h(y)$, an operation which 
is performed\footnote{It may be convenient to separate the 
contributions which originate from the average value $\bh$ 
by redefining $h(y) \rightarrow h(y) -  \bh$.} by using 
eqs.\,(\ref{moments}). This can be graphically represented by 
pairing by means of dotted line the small circles which represent 
the magnetic fields (see for instance Fig.\,1.b), each of which 
gives a factor $\Delta$. These extra terms will certainly increase 
the intricacy of the self--energy of the particle propagator but 
they cannot spoil the integrable nature of the boundary magnetic 
interaction, i.e. the one--to--one nature of the scattering process 
of the particle which hits the boundary. The reason is that the 
graphs which result from the average are similar to those entering 
the pure system which are integrable. Hence, we conclude that the 
dynamics associated with the Ising model with quenched boundary 
magnetic field may also be described by an integrable model. 

This paper is organised as follows: in Section 2 the model is formulated  
in the replica space and the exact reflection $S$--matrix is computed. In 
Section 3 we compute some quenched averaged correlation functions using 
the exact scattering theory. In Section 4 we apply the Thermodynamical 
Bethe Ansatz to compute boundary entropies. Our conclusions are in 
Section 5.   

\resection{Boundary scattering in replica space}

The quenched averages of the disordered systems can be obtained by using 
the so--called replica trick which is based on the identity 
$\overline{\log Z} = \lim_{n\rightarrow 0} (\overline{Z^n} -1)/n$, where 
$\log Z$ is the free energy of the model under investigation. In 
this way the original random problem is mapped onto a quantum field 
theory problem, described by an effective action involving $n$
degrees of freedom in the limit $n \rightarrow 0$. Apart from  
subtleties in taking the $n \rightarrow 0$ limit, the advantage 
of this transformation is that it makes available to us powerful tools 
of quantum field theory and therefore opens up a way of obtaining 
non--perturbative results on the original random system. For the 
problem of interest in this paper, it is easy to see that by using 
(\ref{moments}), the effective action is given by 
\bea
{\cal A}^{(rep)} &\,=\,& \sum_i {\cal A}^{(i)}_0 \,
+ \overline{h}\,\int_{-\infty}^{+\infty} dy \sum_a^n \sigma_a(y) 
\label{replica}
\\
&& \hspace{3mm} - \,\D \int_{-\infty}^{+\infty} dy
\sum_{a < b}^n \sigma_a(y) \, \sigma_b(y) \,\,\,. \nonumber
\eea
The action (\ref{replica}) describes $n$ copies of the Ising model hinged
together at the boundary (Fig. 2). In addition to the coupling to the (mean) 
boundary magnetic field, the new interaction in the action (\ref{replica})
describes a process where a particle with replica index $a$ arrives at 
the boundary and then changes to a particle with another label $b$. 
This elementary process may be repeated along the boundary an arbitrary 
number of times until the particle is finally re-emitted. Since no
production events are involved, the boundary action (\ref{replica}) defines
to a boundary integrable theory. In the following we will assume the 
validity of the above argument irrespective of the actual value of $n$, in
particular also for the values $n < 2$. Consequently, there are two possible 
channels the elastic reflections ${\cal R}_{ab}(\t)$ can go through: 
(1) the first channel is the reflection process described by the amplitude 
$P(\theta)$ where the particle $A^{a}(\theta)$ bounces back by 
keeping its original replica index $a$; (2) the second channel is 
described by the amplitude $Q(\theta)$ where the particle $A^{a}(\theta)$
changes its replica index $a \rightarrow b$ as result of the 
interaction (Fig.\,3). Therefore we write 
\EQ
{\cal R}_{ab}(\t) \,=\,P(\t) \delta_{ab} + Q(\t) (1 - \delta_{ab}) \,\,\,.
\label{matriceR}
\EN
In terms of the geometrical representation of Fig.\,2, the dynamics 
consists in the free propagation of the particle on any one of the 
$n$ sheets, according to its replica label: when the particle hits 
the boundary it can either remain on the same plane (with a probability 
$\mid P(\t)\mid^2$) or redirect its motion to another of the $n-1$ 
planes (with a probability $\mid Q(\t)\mid^2$). 

The two amplitudes $P(\t)$ and $Q(\t)$ satisfy the unitarity conditions   
\EQ
\begin{array}{l}
P(\t) \,P(-\t) + (n-1) \,Q(\t) \,Q(-\t) = 1 \,\, ,
\\
P(\t) \,Q(-\t) + Q(\t) \,P(-\t) + (n-2) \,Q(\t) \,Q(-\t) = 0 \,\, ,
\end{array}
\label{unitarity}
\EN
as well as the cross-unitarity equations 
\EQ
\begin{array}{l}
P\left(\frac{i\pi}{2} - \theta\right) = - P\left(\frac{i\pi}{2} +
\theta\right) \,\, ,\\
Q\left(\frac{i\pi}{2} - \theta\right) = \,\,Q\left(\frac{i\pi}{2} +
\theta\right) \,\,\, .
\end{array}
\label{crossing}
\EN
In the cross-unitarity equations we have chosen $S_{ab} = 1 - 2 \delta_{ab}$
as the $S$--matrix relative to the collision processes of the replica in 
the bulk. The system of equations (\ref{unitarity}) and (\ref{crossing}) 
admits two different classes of solutions depending whether $n \neq 2$ or 
$n=2$. 

For $n \neq 2$, the above crossing equations and the second of the unitarity 
equations may be identically satisfied by the following 
ansatz\footnote{A detailed discussion on the solution of eqs.\,
(\ref{unitarity}) and (\ref{crossing}) for $n \neq 2$ can be found 
in Appendix A.}   
\begin{eqnarray}
&& P(\theta) \,=\,(n-2) \left(\frac{\theta}{i \pi} - 
\frac{1}{2} \right) \, f(\theta) \,\,\, ;\nonumber \\
&& \label{RQ} \\
&& Q(\theta) \,=\,f(\theta) \,\, ,
\nonumber 
\end{eqnarray}
with $f(\t)$ being a crossing symmetric function 
\EQ
f(\theta) \,=\,f(i \pi - \theta) \,\,\, .
\label{ffunction}
\EN
Plugging this ansatz into the first unitarity equation 
we obtain the additional equation 
\EQ
f(\t) \, f(-\t) \,=\, \frac{1}{4 (n-2)^2} \frac{1}
{\left(\frac{\t}{2 \pi}\right)^2 + \left(\frac{n}{4 (n-2)}\right)^2} 
\label{unit}
\EN 
for the function $f(\t)$. The simplest ``minimal'' solution of 
(\ref{ffunction}) and (\ref{unit}) is given by\footnote{The most 
general solution is obtained by multiplying (\ref{solution1}) by 
an arbitrary CDD factor, i.e. an arbitrary meromorphic function $\Phi(\t)$
which satisfies both the equations $\Phi(\t) \Phi(-\t) =1$ and $\Phi(\t) = 
\Phi(i \pi -\t)$. }   
\EQ
f(\t) = \frac{1}{(n-2)}\,\frac{1}{\left(\frac{\t}{i \pi} - \frac{n}{2
(n-2)}\right)} \, F(\t) \,\, ,
\label{solution1}
\EN 
where 
\EQ
F(\t) \,=\, 
\frac{\Gamma\left(\frac{n}{4 (n-2)} + \frac{1}{2} - \frac{\t}{2\pi
i}\right) \,\Gamma\left(\frac{n}{4 (n-2)} + \frac{\t}{2\pi i} \right)} 
{\Gamma\left(\frac{n}{4 (n-2)} + \frac{1}{2} + \frac{\t}{2\pi i}\right) 
\,\Gamma\left(\frac{n}{4 (n-2)} - \frac{\t}{2\pi i} \right)} \,\,\,.
\label{solution2}
\EN 
The function $F(\t)$ admits the following integral representation 
\EQ
F(\t) \,=\,\exp\left[i \int_0^{\infty} \frac{dt}{t} 
\frac{e^{-\frac{t}{2 (n-2)}}}{\cosh\frac{t}{4}} \sin\frac{\t t}{2 \pi}
\right] \,\,\, .
\label{integral}
\EN
Note that the above minimal solution does not contain any free parameters. 
Apart from possible CDD factors, this seems to be a general property of the
system of equations (\ref{unitarity}) and (\ref{crossing}) for generic $n$.
For $n=2$, however, the situation is different. This is due to a simple
reason: in fact, by unfolding two semi-infinite planes the system becomes 
in this case an ordinary Ising model (in the infinite plane) but with a 
line of defect. Hence for $n=2$ the system of equations (\ref{unitarity}) 
and (\ref{crossing}) admits extra solutions, more precisely 
those which belong to the one--parameter family of $S$-matrices of the
Ising model with a line of defect which were determined in \cite{DMS}. In
the present notation they may be written as 
\begin{eqnarray}
&& P(\t) = i \frac{\sin\chi \,\cosh\t}{\sinh\t - i \sin\chi} \,\,\, ;
\nonumber \\
&& \label{defect} \\
&& Q(\t) = \frac{\cos\chi \sinh\t}{\sinh\t - i \sin\chi} \,\, ,
\nonumber 
\end{eqnarray}
where 
\EQ
\sin\chi = - \frac{\Delta}{1 + \frac{\Delta^2}{4}} \,\,\,.
\label{couplingdep}
\EN 
These additional solutions at $n=2$ may shed light on the identification 
of which physical situation the minimal solution (\ref{solution1}) 
corresponds to. Let us compare, in fact, the minimal solutions $P(\t)$ 
and $Q(\t)$ in the limit $n \rightarrow 2$ with the reflection amplitudes 
of the defect line: in this limit $P(\t)$ and $Q(\t)$ go to $0$ and to $-1$
respectively and therefore, from (\ref{defect}) and (\ref{couplingdep}) we
see that they correspond to an $S$-matrix of the defect line with an 
infinite value of the coupling constant $\Delta$. This comparison suggests
that the minimal solution (\ref{solution1}), without any adjustable
parameter, may correspond to $n$ Ising models infinitely coupled at the 
boundary, i.e. to the strongly disordered case $\Delta \rightarrow \infty$. 

Note that for $n=1$ (and $\overline h = 0$) we cannot of course have 
any interaction between the replica and correspondingly the diagonal 
amplitude $P(\t)$ in this case reduces to 
\EQ
P(\t) = -i \coth\left(\frac{i \pi}{4} - \frac{\t}{2}\right) \,\, ,
\label{free}
\EN 
which coincides with the reflection matrix of the pure Ising model 
with free boundary conditions. This result indicates that the simplest 
way to introduce the dependence  on the (mean) magnetic field 
${\overline h}$ is to multiply the minimal solution (\ref{solution1}) 
by the CDD factor  
\EQ
H(\t) \,=\,
\tanh^2 \left(\frac{\t}{2} - \frac{i \pi}{4}\right)
\frac{i \sinh\theta  - \kappa}{i
\sinh\theta + \kappa} \,\,\,. 
\label{CDD}
\EN 
For $n \rightarrow 0$ we have a well--defined limit and the exact
boundary  $S$-matrix is given in this case by 
\begin{eqnarray}
&& P(\t) \,=\, \frac{\t -  \frac{i \pi}{2}}{\t} \,
\frac{\Gamma\left(\frac{1}{2} - \frac{\t}{2 \pi i}\right) 
\Gamma\left(\frac{\t}{2\pi i}\right)}
{\Gamma\left(\frac{1}{2} + \frac{\t}{2 \pi i}\right)
\Gamma\left(-\frac{\t}{2\pi i}\right)} \,H(\t) \,\, ,\nonumber \\
&& \label{n=0} \\
&& Q(\t) \,=\, - \frac{ i \pi}{2 \t} \, 
\frac{\Gamma\left(\frac{1}{2} - \frac{\t}{2 \pi i}\right)
\Gamma\left(\frac{\t}{2\pi i}\right)}
{\Gamma\left(\frac{1}{2} + \frac{\t}{2 \pi i}\right)
\Gamma\left(-\frac{\t}{2\pi i}\right)} \, H(\t) \,\,\, .
\nonumber
\end{eqnarray} 
In the opposite limit $n \rightarrow \infty$ , we find that $Q(\t)$ goes to
zero whereas $P(\t)$ becomes 
\EQ
P(\t) \,=\, - i \tanh^3\left(\frac{i \pi}{4} - \frac{\t}{2}\right) 
\frac{i \sinh\theta  - \kappa}{i
\sinh\theta + \kappa} \,\, , 
\label{ninfinity} 
\EN
i.e. a pure phase.

It is interesting to study the structure of the boundary bound states 
by varying $n$ (in the following we will only consider the case
$\overline h = 0$  since the discussion of the pole structure 
induced by the CDD term $H(\t)$ may be found in \cite{GZ}): 
from eqs.\,(\ref{solution1}) and (\ref{solution2}) we find that the 
positions of the poles are at: 
\begin{eqnarray}
&&\t'_{k} = (2k+1) \, \pi\, i  + \frac{i \pi}{2}
\left(\frac{n}{n-2}\right) \,\,\,;\nonumber
\\ 
&& \label{poles} \\
&&\t''_{l} = - 2 l \,\pi \,i - \frac{i \pi}{2} 
\left(\frac{n}{n-2}\right) \,\, ,
\nonumber
\end{eqnarray}
where $k,l = 0,1,2,...$. For $n > 2$ there are no poles in the physical strip
$0\leq \mbox{\rm Im}\, \t \leq \pi/2$. Vice-versa, for $n < 2$, 
as long as $n$ belongs to the interval $\frac{4k+1}{2k+1}\leq n \leq 
\frac{8k+4}{4k+3}$ (respectively $\frac{8l}{4l+1}\leq n \leq
\frac{4l+1}{2l+1}$), we find that in the physical strip 
there is the unique pole $\t'_{k}$ (respectively $\t''_{l}$).  
These poles are associated to boundary bound states and their existence 
appears compatible with the analysis of the renormalization group 
relative to the variable $\Delta$ \cite{Cardyb,Turban}. In fact, 
for $n > 2$ $\Delta$ is a (marginal) relevant variable, hence of feeble
effect at the short distance scales near the boundary; for $n < 2$, 
$\Delta$ becomes on the contrary a (marginal) irrelevant variable and
therefore it may have quite a strong effect at the short distance scales 
near the boundary: this may result in an attractive force which can be
strong enough to produce bound states. Note that the above poles 
(\ref{poles}) are generically simple poles but for specific values 
of $n$ a pair of them may coincide, giving rise to a double order pole. 
The residue at the simple poles is related to the (square) of the 
boundary--particle couplings (Fig. 4) \cite{GZ}  
\EQ
R_{a}^{b}(\t) \,\sim\, \frac{i}{2} 
\frac{g_{a 0 \alpha} g_{\alpha b 0}} {\theta - i v_{0 a}^{\alpha}} \,\,\,.
\label{residue}
\EN
So, for instance, looking at the pole at $\t = - i\frac{\pi}{2}
\left(\frac{n}{n-2}\right)$  with $ 0 < n < 1$, we have for 
the boundary--particle coupling in the diagonal channel $P(\t)$ 
\EQ
g_{i 0 \alpha}^2 = 2 \sqrt{\pi} \frac{\Gamma\left(\frac{2n-3}{n-2}\right)}
{\Gamma\left(\frac{3 n -4}{2 (n-2)}\right)} \,\,\, .
\label{square}
\EN 
Analogous expressions are found for the residues at the other poles. 

\resection{Quenched Correlation Functions}

Quenched averaged correlation functions of the model may be computed 
by using the replica formalism and quantum field theory methods. 
Here the calculation for the first few is addressed, namely 
the one and two-point functions of the energy and magnetization 
operators. To this aim the following general formulas are needed  
\begin{eqnarray}
&& \overline{\langle \varphi(x) \rangle} \, = \,
\lim_{n\rightarrow 0} \,\langle \varphi_a(x)\rangle \,\,\,;\nonumber \\
&& \overline{\langle \varphi(x) \, \varphi(y) \rangle} \, = \, 
\lim_{n\rightarrow 0} \,\langle \varphi_a(x) \varphi_a(y) \rangle \,\,\,;
\label{replicarule} \\
&& \overline{\langle\varphi(x)\rangle \, \langle \varphi(y) \rangle} 
\,=\, \lim_{n\rightarrow 0} \,\langle \varphi_a(x)
\varphi_b(y)\rangle\mid_{a\neq b} \,\,\,.\nonumber
\end{eqnarray}
In our case these formulas have to be further specified to take 
into account the boundary effects. Let us see how this can be done. 

In the presence of boundary interactions, a convenient way to perform 
the calculation of the correlation functions is to employ a method 
based both on the boundary state wave function and the form 
factors relative to the local operators in the bulk\footnote{For the model
under investigation in this paper, the form factors are all known, being 
the form factors of the thermal Ising model. Their expression may be
found for instance in \cite{Ising}.}: assuming that the boundary is placed
at the  ``time'' $t=0$ and described by its wave function $\mid B>$, 
the correlation functions may be in fact expressed as 
\EQ
\langle {\cal O}_1(x_1,t_1) \ldots {\cal O}_p(x_p,t_p) \rangle  \,=\,
\frac{
\langle 0\mid T_t\left[{\cal O}_1(x_1,t_1) \ldots {\cal
O}_p(x_p,t_p)\right]\mid B \rangle } {\langle 0 \mid B\rangle }
\,\,\,.
\label{correlators} 
\EN 
Since in this geometric setting the Hilbert space of the theory is the
same as that in the bulk, the local operators ${\cal O}_a$ can be
completely characterized by their bulk form factors 
\[
\langle A_{i_1}(\t_1),
\ldots,A_{i_p}(\t_{p})\mid {\cal O}_a(0,0) \mid A_{i_{p+1}}(\t_{p+1}),
\ldots,A_{i_q}(\t_q)\rangle \,\, ,
\] 
irrespectively of the presence of the boundary. Hence, once the 
boundary state wave function is known, the calculation of the correlation
functions (\ref{correlators}) is in principle just a matter of introducing
the completeness relationships of the intermediate states between the 
various operators. In our model the boundary state wave function is 
explicitly given by 
\EQ
\mid B \rangle = \exp\left[\frac{1}{2}\int_{-\infty}^{+\infty} d \theta 
K^{ab} (\theta) A_a(-\theta) A_b(\theta)\right] \mid 0 \rangle \,\, ,
\label{boundarystate}
\EN
where $K^{ab}(\theta) \equiv \hat P(\theta) 
\delta^{ab} + \hat Q(\theta) (1 - \delta^{ab})$, with $\hat
P(\theta) = P\left(\frac{i \pi}{2} - \theta\right)$ and 
$\hat Q(\theta) = Q\left(\frac{i \pi}{2} - \theta\right)$. 
With the above information, let us proceed to the calculation of 
some correlation functions. 

The simplest one is the averaged one--point function of the energy
operator: in the language of the replica this is given by  
\EQ
\epsilon_0(t) \equiv \overline{\langle\epsilon(\rho)\rangle}\, 
= \,\lim_{n\rightarrow 0} \, 
\langle 0\mid \epsilon_a(\rho)\mid B \rangle \,\, ,
\label{energy}
\EN
where $\rho \equiv (x,t)$ and $t$ is the distance from the boundary. 
Once averaged, the one-point function $\overline{\langle\epsilon(t)\rangle}$ 
does not depend on $x$, as a consequence of the translation invariance 
along the boundary. In the bulk, the operator $\epsilon_a(x,t)$ couples 
only to the two--particle state with the same index $a$ and its exact 
form factor is given by \cite{Ising}
\begin{eqnarray}
&& \langle 0\mid \epsilon_a(x,t)\mid A_a(\t_1) A_a(\t_2)\rangle \,=\,
2 \pi m \, i\,\, \sinh\frac{\t_1-\t_2}{2}  
\label{FF}
\\
&& \hspace{30mm} \times \,\exp\left[-m t\, (\cosh\t_1 + \cosh\t_2) +
i m x \,(\sinh\t_1 + \sinh\t_2)\right]\,\, .
\nonumber 
\end{eqnarray}
Hence there is only one term of the boundary state wave function 
which contributes in this case (see Fig.\,5) and we have  
\EQ
\overline{\langle\epsilon(t)\rangle} = 
\ -i \, m \,\int_0^{\infty} d\theta
\sinh\theta \,\hat P(\theta) \,e^{-2mt\cosh\theta} \,\,\, .
\label{e-profile}
\EN
In the above formula, $\hat P(\t)$ is understood to be 
evaluated\footnote{Unless explicitly stated, from now on both $\hat P(\t)$ 
and $\hat Q(\t)$ are meant to be the functions evaluated at $n=0$.}  
at $n=0$. Various profiles of this correlation function for different values 
of the mean magnetic field $\overline h$ may be found in Fig.\,6. 
They present the typical cross--over associated to a relevant 
boundary operator: for $\overline h =0$ the correlator present a 
short--distance singularity $\overline{\langle\epsilon(t)\rangle} 
\sim - 1/(2 t)$ whereas for $\overline h = \infty $ there is a swap 
of the overall sign $\overline{\langle\epsilon(t)\rangle} \sim 1/(2 t)$. 
For finite values of $\overline h$ (however large), the corresponding 
curves follow the behavior of the curve at $\overline h =0$ at very short 
distances while they tend to follow the behavior of the curve relative to 
$\overline h = \infty$ at large distances. 

Let us consider now 
\EQ
{\cal G}(\rho_1,\rho_2) \,=\, 
\overline{\langle \epsilon(\rho_1) \,
\epsilon(\rho_2) \rangle} \,=\, 
\lim_{n\rightarrow 0} \, \langle 0\mid \epsilon_a(\rho_1) 
\epsilon_a(\rho_2) \mid B \rangle  \,\,\, .
\label{energy2}
\EN
In terms of the replica, there are four possible graphs entering 
the above correlation functions (see Fig.\,7) and correspondingly 
we have  
\EQ
{\cal G}(\rho_1,\rho_2)\,=\,I_1 + I_2 + I_3 + I_4\,\,\, .
\EN
The first of them (Fig.\,7.a) is just the product of the vacuum 
expectation values,  
\EQ
I_1 = 
\epsilon_0(t_1)
\epsilon_0(t_2)  \,\,\, .
\label{vacuump}
\EN  
Let us adopt in the following the notation $\overline \tau\equiv t_1 + t_2$, 
$\tau \equiv t_1-t_2$, $x\equiv x_1 - x_2$,  $r \equiv 
\sqrt{x^2 + \tau^2}$. The second term (see Fig.\,7.b) is the bulk 
energy-energy correlation
function, which, in terms of the Bessel function $K_0$ may be written
as  
\EQ
I_2 \, = \, \left[\left(\frac{\partial }{\partial x} 
K_0(mr)\right)^2 + \left(\frac{\partial}{\partial \tau} K_0(mr)\right)^2 
- m^2 \,(K_0(mr))^2 \right]\,\,\,.
\EN 
The remaining two terms relative to the graphs of Figs.\,7c. and 7.d 
involve the amplitude $\hat P(\t)$ and their expressions may be cast 
in the form 
\begin{eqnarray}
I_3 &=& \,\left[2 m 
\left(\frac{\partial}{\partial x} K_0(m r) \right) 
F(x,\overline \tau) - 2 m  
K_0(m r) \left(\frac{\partial}{\partial x} F(x,\overline \tau)\right) 
\right]\,\,\,;  \nonumber \\
&& \label{234} \\
I_4 &=&  
\left[\left(m F(x,\overline \tau)\right)^2 - 
\left(\frac{\partial}{\partial x} F(x,\overline \tau)\right)^2 - \,
\left(\frac{\partial}{ \partial \overline \tau} F(x,\overline \tau)
\right)^2 \right] \,\, ,\nonumber  
\end{eqnarray}
where we have introduced the auxiliary function 
\EQ
F(x,\tau)\,=\, \frac{1}{2}\,\int_{-\infty}^{+\infty} d\beta \,\hat P(\beta) 
\exp[-m \tau \cosh\beta + i m x \sinh\beta]
\,\,\,.
\label{auxi1}
\EN 
Altogether, the two-point function (\ref{energy2}) can be finally 
expressed as 
\begin{eqnarray}
{\cal G}(\rho_1,\rho_2) &=& \epsilon_0(t_1) \epsilon_0(t_2) + 
\left[\frac{\partial}{\partial x} K_0(m r) +   
m F(x,\overline \tau)\right]^2 + \left[\frac{\partial}{\partial \tau} 
K_0(m r)
\right]^2 \nonumber \\ 
& & - \left[\frac{\partial}{\partial\, \overline \tau} 
F(x,\overline \tau)\right]^2 - \left[m K_0(m r) +  \frac{\partial}
{\partial x} F(x,\overline \tau)\right]^2 
\,\,\,. \label{fside}
\end{eqnarray}
In the case of the energy operator, it is also interesting to
calculate the following quenched averaged correlation function 
\EQ
{\cal G}_A(\rho_1,\rho_2) \,=\,
\overline{\langle\epsilon(t_1)\rangle   
\langle\epsilon(t_2)\rangle}
\,=\, \lim_{n\rightarrow 0} \, \langle \epsilon_a(x)
\epsilon_b(x)\rangle\mid_{a\neq b} \,\,\,.
\label{energy2ea}
\EN 
There are only two graphs which contribute to this correlation function.  
The first one is the disconnected term given by the product of the 
vacuum expectation values, eq.\,(\ref{vacuump}). The second one is 
the graph drawn in Fig.\,8 which involves the off--diagonal amplitude $\hat
Q(\t)$. The correlator ${\cal G}_A(\rho_1,\rho_2)$ can be expressed 
in this case as 
\begin{eqnarray}
{\cal G}_A(\rho_1,\rho_2) & = & 
\,\left[\left(\frac{\partial}{\partial x } Z(x,\overline \tau)\right)^2 + 
\left(\frac{\partial}{\partial \bar \tau} Z(x,\bar \tau)\right)^2 - 
\left(m Z(x,\bar \tau)\right)^2\right] + 
\nonumber \\
&& \hspace{1mm} + \,\epsilon_0(t_1) \,\epsilon_0(t_2) 
\,\, , 
\label{facross} 
\end{eqnarray}
where we have defined the auxiliary function   
\EQ
Z(x,t)\,\equiv \,\frac{1}{2}\int_{-\infty}^{+\infty} d\beta \,\hat Q(\beta)
\exp[-m t \cosh\beta + i m x \sinh \beta] 
\,\,\,.
\label{aux2}
\EN
In the limit $\xi \rightarrow 0$ the composite operator 
$E_{ab}(\rho) = \lim_{\xi \rightarrow 0} \epsilon_a(\rho + \xi)
\epsilon_b(\rho)$ does not present ultraviolet singularities: 
in fact, the two fields $\epsilon_a$ and $\epsilon_b$ do not 
have short--distance divergences in the bulk and they are 
linked to each other only through the boundary interaction. 
Therefore the above formula (\ref{facross}) can be specialized 
to the case $\rho_1=\rho_2$ and studied correspondingly as 
one-point function of the field $E_{ab}(t)$: its profile versus 
the distance from the boundary is plotted in Fig.\,9 for different 
values of the mean magnetic field. It is interesting to note the 
rapid crossover which occurs in this correlation function for large 
values of $\overline h$ at very short distance scales.  

Let us now turn our attention to some correlation functions of 
the magnetization operators. The simplest is the one--point 
function of the disorder operator $\mu(\rho)$
\EQ
\mu_0(t)\,=\, \overline{\langle\mu(\rho)\rangle}\, 
= \,\lim_{n\rightarrow 0} 
\langle 0\mid \mu_a(\rho)\mid B \rangle \,\,\,.
\label{mu}
\EN
In the high--temperature phase, the disorder operator $\mu_a(\rho)$ 
has a non-zero vacuum expectation value and couples to all states with an
even number of particles with the same replica index: its 
explicit form factors are given by\footnote{The magnetization field 
defined by the Form Factors normalized as in eq.\,(\ref{FFmu}) differs 
from the corresponding conformal field only by a normalization
constant,  $\mu(\rho) = {\cal F}\, \mu_{conf}(\rho)$, where 
${\cal F} = 2^{-1/12} e^{1/8} A^{-3/2} m^{-1/8}$ and $A = 1.282427$ 
(Glasher constant).} \cite{Ising} 
\EQ
\langle 0\mid \mu_a(0,0)\mid A_a(\t_1) \ldots A_a(\t_{2n})\rangle  
\,= \, (-i)^n \prod_{i<j} \tanh\frac{\t_i-\t_j}{2}
\,\,\, .
\label{FFmu}
\EN
Since the boundary state consists of a condensate of Cooper pairs, i.e. 
couples of particles with equal and opposite momentum, we have 
to specialize the above formula to the case  
$ \langle 0\mid\mu_a (0)\mid A_a(-\t_1) A_a(\t_1) \ldots A_a(-\t_n) 
A_a(\t_n)\rangle$. This matrix element can be conveniently written as 
\EQ
\langle 0\mid \mu_a(0,0)\mid A_a(-\t_1) A_a(\t_1),\ldots A_a(-\t_n) 
A_a(\t_n) \rangle \,=\,
i^n \left(\prod_{i=1}^n \tanh\t_i\right) 
\times \det\, W(\t_i,\t_j)
\,\, ,
\EN
where $W(\t_i,\t_j)$ is an $n\times n$ matrix whose elements are given by 
\EQ
W(\t_i,\t_j)\,=\, \left(\frac{2\,\sqrt{\cosh\t_i\cosh\t_j}}
{\cosh\t_i+\cosh\t_j}
\right)\,\,\,.
\EN
The one--point function of the disorder operator is made of an infinite 
number of terms, as the one drawn in Fig.\,10. Its final expression can 
be conveniently written as a Fredholm determinant  
\begin{eqnarray}
\overline{\langle\mu(t)\rangle} & = & \sum_{n=0}^{\infty} \frac{1}{n!}
\int_{-\infty}^{\infty}
d\t_1\ldots d\t_n \,z^n\,\left(\prod_{k=0}^n i\,\tanh\t_k \,
\hat P(\t_k)\,e^{-2mt \cosh\t_k} \right)
 \,\det \,W(\t_i,\t_j) \,= \nonumber \\
& & =\, \det\, (1 + z\,{\cal W}) \,\, ,
\label{fredholm}
\end{eqnarray}
where the kernel of the integral operator is given by 
\EQ
{\cal W}(\t_i,\t_j)\,=\,\frac{E(\t_i, mt) 
E(\t_j,mt)}{\cosh\t_i + \cosh\t_j}\,\, ,
\label{kernel}
\EN
and 
\EQ
E(\t,mt) \,= \,e^{-mt \cosh\t}\, \left(i\hat P(\t)\,
\tanh\t \right)^{1/2} \hspace{3mm},\hspace{5mm}
z \,=\,\frac{1}{2\pi} \,\,\, .
\EN 
As consequence of the translation invariance along the boundary, 
$\mu_0(t)$ depends only on its distance $t$ from the boundary. 
In terms of the eigenvalues of the integral operator and their multiplicity, 
$\overline{\mu(t)}$ can be also expressed as 
\EQ
\overline{\mu(t)}\,=\,\prod_{i=1}^{\infty} \left(1 +
z\,\lambda_i\right)^{a_i} \label{eigenvalues} \,\,\,.
\EN
As far as $mt$ is finite, the kernel is square integrable and therefore 
its properties are those of a bounded symmetric integral operator 
\cite{Integral}. For large values of $mt$, $\overline{\langle\mu(t)\rangle}$ 
falls off exponentially to its bulk vacuum expectation value. However, when
$mt \rightarrow 0$, the integral operator becomes unbounded: in this case, 
the multiplicity of the eigenvalues grows logarithmically as 
$a(t) \sim \frac{1}{\pi} \ln\left(\frac{1}{mt}\right)$ and the one-point 
function presents a power law behavior $\overline{\langle\mu(t)\rangle}
\sim A/(2 t)^{\zeta}$. To determine $\zeta$, observe that as long as 
$n \neq 2$ (which we always assume to be the case in the following) and
$\overline h \neq \infty$, we have $\lim_{\t\rightarrow \infty} \hat P(\t)
= -i$. Let us denote this limit as $\hat P_{-}$. For $n\neq 2$ and 
$\overline h = \infty$, we have $\lim_{\t\rightarrow\infty} \hat P(\t) = i$, 
instead. This limit value will be denoted by $\hat P_{+}$. To study the 
short distance behavior $m t \rightarrow 0$, we can reasonably substitute 
$\hat P(\t)$ with its asymptotic limits $\hat P_{\pm}$. The eigenvalues of
the integral operator then become dense in the interval $(0,\infty)$ 
according to the distribution 
\[
\lambda(p) \,=\,\frac{2\pi}{\cosh\pi p}
\]
and for the exponent $\zeta_{\pm}$ relative to the two cases we find 
\EQ
\zeta_{\pm} \,= -\,\frac{1}{\pi}\,
\int_0^{\infty} dp \,\ln\left(1 \pm \frac{2\pi z}{\cosh p}\right)
\,=\,- \frac{1}{8} +\frac{1}{2\pi^2} \arccos^2(\mp 1) = 
\left\{\begin{array}{c}
\,\,\,\frac{3}{8} \\
-\frac{1}{8} 
\end{array}
\right.
\label{surfaceexp}
\EN 
The profile of this one-point function is drawn in Fig.\,11 for 
several values of $\overline h$: the observed cross-over behavior 
of the curves by varying $\overline h$ is perfectly analogous to 
the one of the energy operator. 

Let us conclude this section with the discussion relative to the
(averaged) two-point function of the magnetization operator 
$\sigma(\rho)$. In the bulk, this operator has non--vanishing matrix 
elements on an odd number of particles and their expression is given 
by the same formula as (\ref{FFmu}) \cite{Ising}. Consider first 
\EQ
{\cal G}^{\sigma}(\rho_1,\rho_2) \,=\,
\overline{\langle\sigma(\rho_1) 
\sigma(\rho_2)\rangle} = \lim_{n\rightarrow 0}\, 
\langle 0\mid\sigma_a(\rho_1) \sigma_a(\rho_2) \mid B \rangle \,\,\,. 
\label{first}
\EN 
There are two kinds of graphs entering the above correlator (see 
Fig.\,12). The first does not involve the boundary and therefore 
the sum of such graphs gives rise to the known correlation function in the
bulk which is expressible in terms of Painleve' function \cite{McCoy}. The
second set of graphs is made of an arbitrary number of pairs of particles
$A_a$ emitted by the boundary and absorbed by the two operators. 
The two operators, in turn, are also linked together by an arbitrary 
(odd) number of intermediate particle states. The first non-trivial
contribution to this correlation function is the one shown in Fig.\,13, 
given by 
\EQ
J(\rho_1,\rho_2) \,=\, 2g^2 F(x,\bar \tau) \,\, ,
\label{apr1}
\EN 
where $g$ denotes the (constant) one-particle form factor of the
magnetization operator $\langle 0 \mid  \sigma(0)_a \mid  A_b(\b) \rangle
= g \delta_{ab}$ and the function $F(x,t)$ is defined in (\ref{auxi1}). 
A convenient form of the series resulting from the graphs of Fig.\,12  
is presently unknown.  

Let us consider now the other type of two--point averaged correlation 
function
\EQ
{\cal G}^{\sigma}_A(\rho_1,\rho_2) \,=\,\overline{\langle\sigma(\rho_1) 
\rangle \,\langle \sigma(\rho_2)\rangle} = \lim_{n\rightarrow 0}\, 
\langle \sigma_a(\rho_1) \sigma_b(\rho_2) \rangle \mid_{a \neq b} \,\,\,.
\label{second}
\EN 
The graphs contributing to this correlation function are those shown 
in Fig.\,14. This time the boundary can emit an arbitrary number 
of pairs of particles of type $A_a A_a$, $A_b A_b$ or $A_a A_b$: the 
former two pairs are individually absorbed by each operator whereas the 
latter link the operators to each other. The simplest (non--disconnected)
term of the resulting series is given in this case by  
\EQ
J_A^{(1)}(\rho_1,\rho_2) \,=\, 2g^2 Z(x,\bar t)\,\, ,
\EN
where the function $Z(x,\bar t)$ is defined in eq.\,(\ref{aux2}). 
The above discussion can be easily generalized to the correlator 
${\cal G}^{\mu}_A(\rho_1,\rho_2) = \overline{\langle\mu(\rho_1)\rangle 
\langle \mu(\rho_2)\rangle}$ as well. In this case the lowest 
contribution of the series originates from a graph like the one 
of Fig.\,8 and its explicit expression is given by   
\begin{eqnarray}
J_A^{(2)}(\rho_1,\rho_2) \, = \, \int_{0}^{+\infty}
\frac{d\b_1}{2 \pi} \int_{-\infty}^{+\infty}
\frac{d\b_2}{2 \pi} \, \tanh^2(\frac{\b_1-\b_2}{2})\, 
\hat Q(\b_1) \,\hat
Q(\b_2) \times \nonumber \\ e^{-m(t_1+t_2)(\cosh\b_1+\cosh\b_2)}
\cos[m(x_1-x_2)(\sinh \b_1+\sinh \b_2)] \,\,\,.
\label{apr1A}
\end{eqnarray}
It can be rewritten as
\begin{eqnarray}
J_A^{(2)}(\rho_1,\rho_2) \,=\, 4 \sum_{l=0}^{+\infty}\left[ 
(2l+1)\, \left({\cal Z}_{2l+2}(x,\bar \tau) {\cal Z}_{2l} 
(x,\bar \tau) - 
{\cal Z}_{2l+1}^2(x, \bar \tau) \right) + 
\label{calsberg} \right.\\
\\
\left.+ (2l+2)\, 
\left({\cal Z}_{2l+3}(x,\bar \tau) {\cal Z}_{2l+1}
(x,\bar \tau) - {\cal Z}_{2l+2}^2(x,\bar \tau)\right)
\nonumber
\right] 
\, , 
\end{eqnarray}
where  
\EQ
{\cal Z}_l(x,t)\,\equiv \,
\frac{1}{2}\int_{-\infty}^{+\infty}\frac{d\b}{2\pi}\left(\tanh
\frac{\b}{2}\right)^l\, \hat Q(\b) \exp[-mt\cosh\b+imx\sinh\b] \, .
\label{fl}
\EN
Since the rate of convergence of the functions ${\cal Z}_l(x,t)$ 
becomes faster and faster by increasing $l$, the series (\ref{calsberg}) 
can be efficiently estimate by means of the first few terms. 

As was the case for the energy operator, we can define the 
operators
\EQ
\Sigma_{ab}(\rho) = 
\lim_{\xi \rightarrow 0} 
\sigma_a(\rho + \xi) \sigma_b(\rho) 
\,\,\,\, ; \,\,\,
\Upsilon_{ab}(\rho) = \lim_{\xi \rightarrow 0} 
\mu_a(\rho + \xi) \mu_b(\rho)\,\,\, ,
\label{EdAn} 
\EN
without facing any short--distance singularities. It is therefore 
possible to study the correlator ${\cal G}^{\sigma,\mu}_A(\rho_1,\rho_2)$ 
in the limit $\rho_1 \rightarrow \rho_2$, the so-called Edwards-Anderson
(EA) order parameters $\overline{\langle\sigma(\rho)\rangle^2}$ and 
$\overline{\langle\mu(\rho)\rangle^2}$. The profile of the Edwards-Anderson 
order parameter $\overline{\langle\mu(\rho)\rangle^2}$ for different 
values of $\bar h$ by using the lowest order approximation (\ref{apr1A}) 
is drawn in Fig.\,15. Presently however it is quite difficult to make any
comparison with the lowest order expression in $\Delta$ ($\bar h =0$) 
for such parameter   
\EQ
\overline{\langle\mu(t)\rangle^2} \sim 
\Delta \frac{1}{\,\,r^{1/4}} \,\, ,
\label{cardyapr}
\EN
which was obtained by Cardy \cite{Cardyb}. This difficulty is due to 
several reasons: (a) the different methods which were employed in the 
two cases (Cardy obtained in fact (\ref{cardyapr}) by using boundary
conformal perturbation theory; on the other hand, our approach is based 
on the scattering theory relative to the off--critical excitations); (b) 
the lack of an exact re-summation of the series originated from the
graphs of Fig.\,14 which, if known, would permit in principle a 
comparison between the two methods above and finally (c) that the result 
(\ref{cardyapr}) applies for small values of $\Delta$ whereas our scattering 
theory seems to apply to the opposite limit $\Delta\rightarrow \infty$
instead. 

\resection{Boundary Thermodynamical Bethe Ansatz}

Once the exact $S$--matrix of a model is known, finite--size effects and 
associated thermodynamical quantities can be calculated by means of the 
Thermodynamical Bethe Ansatz (TBA): the case of systems with relativistic
invariance and a cylinder geometry with periodic boundary conditions 
has been put forward in \cite{TBAZ}. What is of interest for us here is 
the generalization of the TBA to systems with boundary: such
generalization can be found discussed in \cite{TBAb}. 
 
Consider then an Ising model defined on a cylinder of width $L$ and length
$R$ with boundary conditions at its extremities described by the reflection
scattering theory of Sect.\,2 (see Fig.\,16.a). Choosing as direction of
the time the horizontal axis between the two boundaries, the partition 
function of the model can be expressed as  
\EQ
Z = \langle B \mid \exp(-R H)\mid B \rangle \,\,\, ,
\label{partition}
\EN 
where $H$ is the Hamiltonian of the bulk system with periodic boundary
conditions. In the limit $m R \gg 1$, the partition function reduces to 
\EQ
Z \sim g_I g_{II} \,e^{- R E_0} 
\,\,\, ,
\label{entropy}
\EN 
where $E_0$ is the ground--state energy of the system whereas 
$g_I$ and $g_{II}$ are the boundary degeneracy at each end of the 
cylinder. The $R$--independent term  
\EQ
{\cal S}_{I,II} \,=\, \ln g_{I,II}
\EN
can be then interpreted as boundary entropies of the system \cite{AL}. 
They can be computed by using the boundary Thermodynamical 
Bethe Ansatz \cite{TBAb}. The application of this method 
requires first of all to look at cylinder geometry of Fig.\,16.a in 
a different way, namely the time evolution should be regarded as it takes
place along the vertical axes, i.e. along the circumference of length $L$. 
Hence, the dynamics consists in a set of $N$ particles which move and scatter 
each other along the segment of length $R$ until they reach the
boundaries: then, they are reflected off with amplitudes
${\cal R}_{ab}(\t)$. The finite geometry of the system induces a quantization
condition on the momenta of the particles which can be obtained as follows.
Let $n_a$ be the number of particles of replica index $a$, with $\sum_a
n_a = N$. Consider now one of $N$ particles, of species $b$ and send it 
on a ``round trip'' along the cylinder (Fig.\,16.b): after coming back to
the same location, the wave function of the $N$ particles has picked up a
new phase shift. In fact, 
each time that the particle scatters off a like particle it picks up a $-1$
from their bulk S-matrix (all other scattering processes in the bulk with 
particles of different species do not contribute since the $S$--matrix in
these cases is simply $+1$); in addition to these processes, there is 
an additional phase $R_{bk}(\t) R_{kb}(\t)$ (sum on the internal index
$k$) due to the scattering off the two boundaries. Since the total 
wave function of the $N$ particle state is assumed to be periodic 
under this ``round trip'' along the cylinder, we get the 
quantization condition 
\EQ
e^{2 i R m_b \sinh(\theta)} (-1)^{n_b + n_k-1} (\Lambda_a (\theta))^2 = 1
\ , \ \label{tba1}
\EN
where $\Lambda_a(\theta)$, $a=1,2,\ldots,n$, are eigenvalues of the 
reflection matrix ${\cal R}_{bk}(\theta)$. They can be found explicitly 
since any $n \times n$ matrix of the form 
$$
M_{ij} = (P-Q) \delta_{ij} + Q
$$ 
has only two types of eigenvalues: $\lambda_1=P+(n-1)Q$ with multiplicity 
one and $\lambda_2=P-Q$, which is $(n-1)$ degenerate. 

With this information, the analysis that follows is quite standard 
(see for instance \cite{TBAb}) and therefore we only sketch the main 
steps here. Eq.\,(\ref{tba1}) puts a constraints among the physical
rapidities and therefore for each particle in the interval $\theta$ 
and $\theta+d\theta$ we can define the density of
holes, $\rho^h_i(\theta)$, and of actually occupied states,
$\rho_i(\theta)$. Hence, the density of available states for particles of
species $a$ is $\rho_a(\theta) + \rho_a^h(\theta)$ which is given by 
\EQ
2\pi(\rho_a(\theta) + \rho_a^h(\theta)) = m_a\cosh(\theta)+
{1 \over {2R}} \Phi_a(\theta) \ , \ 
\label{tba2}
\EN 
where we have introduced
\EQ
\Phi_a(\theta) = - 2 i {d  \over {d \theta}} \ln \Lambda_a(\theta)
-2\pi \delta(\theta)
\ . \ \label{tba3}
\EN
The delta function term is introduced in order to remove the unwanted
solution $\theta=0$. The partition function is obtained by expressing 
the Helmholtz's free energy as a functional of the density of states 
(holes and particles) and minimize it. The final expression of the 
boundary entropy is given by the massless limit of
\EQ
{\cal S}_b= \lim_{m \rightarrow 0} \sum_{i=1}^n \int {{d\theta} \over {4
\pi}}  \Phi_i(\theta) \ln(1+e^{-\epsilon_i(\theta)})
\ , \ \label{tba4}
\EN 
where ${\cal S}_b$ is the boundary entropy at one end and the
pseudo--energies $\epsilon_i(\theta)$ are given by the simple 
expression
\EQ
\epsilon_i(\theta)= m_i R \cosh(\theta) \ . \
\EN
since the bulk S-matrix is $ \pm 1$.

The formula (\ref{tba4}) becomes then 
\EQ
{\cal S}_b={n \over \pi}  \int_0^{\infty} dx {1 \over {1+x^2}} \ln
(1+e^{-2\pi a x}) \ , \ \label{tba5}
\EN 
where $m R \cosh(\theta) \equiv 2 \pi a x$ and 
$a = {\overline h}^2 R/(4 \pi)$. Note that each species gives exactly the
same contribution and so we have an overall factor of $n$ in our
expression. The above integral can be computed exactly  
\EQ
{\cal S}_b = n \ln \left({{\sqrt{2 \pi}} \over {\Gamma(a+{1 \over 2})}} 
\left({a \over e} \right)^a \right) \ . \ 
\label{tba6}
\EN
For $n=1$ (which corresponds to the pure Ising model) we obtain the 
correct boundary entropy difference $\Delta {\cal S}_b = \ln \sqrt{2}$ (for
$a=\infty$ and $a = 0$) \cite{AL}. The entropy difference of
the disordered model ($n=0$) is computed by taking the derivative of the 
ground state degeneracy with respect to $n$ and let $n \rightarrow 0$. 
This gives $\Delta {\cal S}_b= \ln \sqrt 2$, exactly the same result as in
the pure Ising model.

\resection{Conclusions} 

In this paper we have proposed an exact scattering theory in the 
replica space of $n$ species of particles $A_a$ (in the limit 
$n \rightarrow 0$) to describe the dynamics of the two--dimensional Ising
model coupled to a boundary random magnetic field. By using methods
borrowed from boundary quantum field theories, we have then computed 
averaged correlation functions of the order parameters as well as 
finite--size effects of the system defined on a strip. It would be 
obviously interesting to test the feasibility of this approach 
for analyzing other models with random interactions localized on 
the boundary.

\vspace{15mm} 
\noindent
{\em Acknowledgments.} We wish to thank J. Cardy, S. Franz, S. Guruswamy 
and R. Monasson for useful conversations and comments. We are extremely 
grateful to J.M. Maillard for helpful discussions. One of us (GM) would
like to thank the organizers of the workshops at the H. Poincare' 
Institute in Paris and the Institute of Theoretical Physics in Santa Barbara 
for the warm hospitality during the staying at these institutes, where part
of this work has been done. (MM) would like to thank S. Randjbar--Daemi 
and the High--Energy group at ICTP for the warm hospitality.

\vspace{15mm}
\noindent
{\em Note Added}. We become aware of the reference \cite{Turban} after 
finishing the present work. We thank L. Turban for his observation. 

Related work, on the 2-dimensional Dirac theory in semi-infinite space with
random boundary interactions, has recently been done independently by S.
Guruswamy and A.W.W. Ludwig \cite{GL}.  
\newpage

\appendix

\appsection

In this appendix we briefly discuss the most general solution of the 
boundary unitarity and cross-unitarity equations (\ref{unitarity}) 
and (\ref{crossing}) for $n \neq 2$. Let's initially define a function
$D(\theta) = P(\theta) - Q(\theta)$. It's easy to see that the unitarity
equations impose
\EQ
D(\t) D(-\t) \,=\, 1 \ . \ \label{ap3}
\EN
Therefore, substituting 
$
P(\t) = D(\t) + Q(\t) 
$
into the first unitarity equation (\ref{unitarity}), we find 
\EQ
D(\t) Q(-\t ) + D(-\t) Q(\t ) +n Q(\t) Q(-\t) = 0 \,\,\,.
\EN
Dividing by $Q(\t) Q(-\t)$, we have 
\EQ
{{D(\t)} \over {Q(\t)}} + {{D(-\t)} \over {Q(-\t)}} + n = 0 \ . \ \label{ap5}
\EN
Define $T(\t)$ by
\EQ
T(\t) = {{D(\t)} \over {Q(\t)}} + { n \over 2} \ . \
\EN
{}From (\ref{ap5}) we see that $T(\t)$ is an odd function. Cross-unitarity 
requires
\EQ
T(i\pi - \t ) + T(\t) = n-2 \ . \  \label{cu}
\EN
With the position $D(\theta) = d(\theta) (2 T(\theta) - n)$, we see that
$d(\theta)$ has to satisfy
\EQ
d(\t) d(-\t) ={1 \over {(2T(\t) - n)(2T(-\t) - n)}} \ . \
\EN
{}From this point on the analysis is standard and one can easily find two 
solutions for $d(\theta)$ (up to CDD factors) 
\EQ
d(\t) \, = \, {1 \over {(2T(\t) - n)}}
{{\Gamma(1-{{n-2T(\t)} \over{4(n-2)}})
\Gamma({1 \over 2}-{{n+2T(\t)}
\over{4(n-2)}})} \over {\Gamma(1-{{{n+2T(\t)}}\over{4(n-2)}})
\Gamma({1\over 2} - {{n-2T(\t)}
\over{4(n-2)}})}} \ . \ \label{ap26}
\EN
\EQ
d(\t) \, = \, {- 1 \over {(2T(\t) + n)}}
{{\Gamma(1+{{n+2T(\t)} \over{4(n-2)}})
\Gamma({1 \over 2}+{{n-2T(\t)}
\over{4(n-2)}})} \over {\Gamma(1+{{{n-2T(\t)}}\over{4(n-2)}})
\Gamma({1\over 2} + {{n + 2T(\t)}
\over{4(n-2)}})}} \ . \ \label{ap7}
\EN
These are general expressions that solve unitarity and cross-unitarity 
conditions provided $T(\t)$ is an odd function satisfying 
the equation (\ref{cu}). The simplest solution for $T(\t)$
which gives the correct analytic behavior is
\EQ
T(\t) = (n-2) {\t \over {i \pi}}\ . \
\EN
With this choice of $T(\t)$, the two solutions equations (\ref{ap26}) and 
(\ref{ap7}) are easily seen to be connected by a CDD factor.
They differ in the positions of the poles: the first one has poles 
in the physical strip for $ n > 2$, the second one, instead, for $ n < 2$.
This last feature is one of the reasons to use the latter solution 
for the physical scattering amplitudes in the replica space.

\newpage

\newpage

{\bf Figure Caption}

\vspace{5mm}

\begin{description}
\item [Figure 1]. Graphs of the perturbative series: (a) pure system 
(b) system with disorder. 
\item [Figure 2]. Replica space. 
\item [Figure 3]. Reflection amplitudes in replica space. 
\item [Figure 4]. Boundary bound state and boundary--particle couplings. 
\item [Figure 5]. Graph of the averaged one--point function of the energy
operator. 
\item [Figure 6]. Profiles of the one--point function of the 
energy operator versus its distance from the boundary for different 
values of 
$\kappa$: $\kappa = 1$ (long dashes), $\kappa = -1
$ (short dashes) and $\kappa = -10 $ (full line). 
\item [Figure 7]. All possible graphs of the averaged two--point function 
of the energy operator. 
\item [Figure 8]. Diagram relative to $\overline{\langle \epsilon(\rho_1) \,
\epsilon(\rho_2) \rangle}$.  
\item [Figure 9]. Profiles of the one--point function 
$\overline{\langle \epsilon(t)\rangle^2}$ for different 
values of $\kappa$: $\kappa = 1$ (long dashes), $\kappa = 
-1 $ (short dashes) and $\kappa = -10 $ (full line). 
\item [Figure 10]. One of the diagrams entering the one--point function 
of the disorder operator. 
\item [Figure 11]. Profiles of the one--point function 
$\overline{\langle \mu(t)\rangle}$ at the lowest order approximation for
different values of $\kappa$: $\kappa = 1$ (long dashes), 
$\kappa = -1$ (short dashes) and $\kappa = -10$ (full line). 
\item [Figure 12]. Diagrams entering the averaged two--point function 
of the magnetization operator. 
\item [Figure 13]. Simplest graph contributing to the averaged 
two--point function of the magnetization operator. 
\item [Figure 14]. Diagrams contributing to the two--point function 
$\overline{\langle \mu(\rho_1)\rangle \,\langle \,\mu(\rho_2) \rangle}$.  
\item [Figure 15]. Profiles of the EA parameter $\overline{\langle
\mu(t)\rangle^2}$ at the lowest order approximation for different 
values of $\kappa$: $\kappa = 1$ (long dashes),
$\kappa = -1$ (short dashes) and $\kappa = -10$ (full line).
\item [Figure 16]. Strip geometry (a) and ``round trip'' (b) of the 
TBA quantization equation. 
\end{description}

\pagestyle{empty}

\newpage
\begin{figure}
\centerline{
\psfig{figure=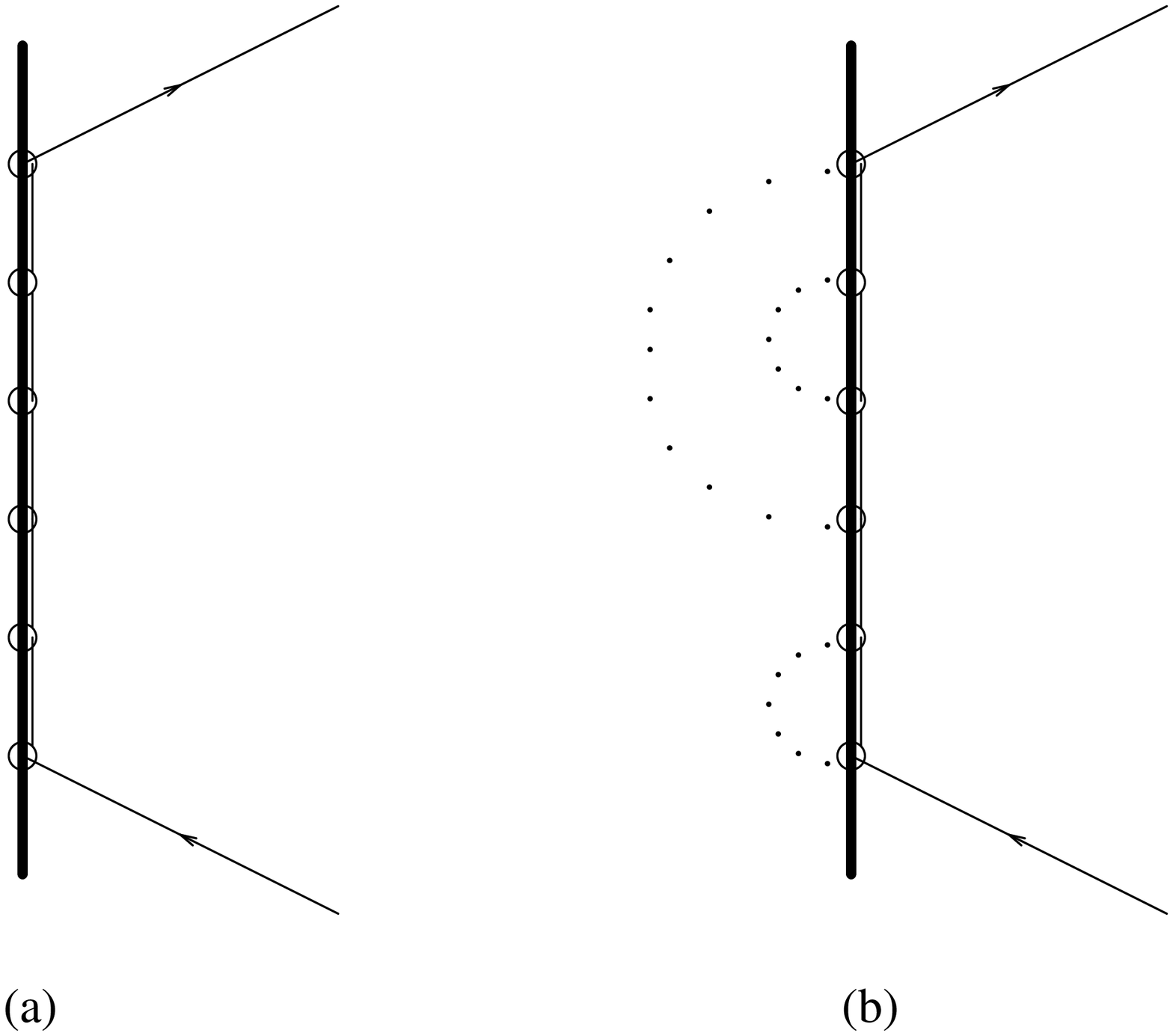}}
\vspace{3cm}
\begin{center}
{\bf Figure 1}
\end{center}
\end{figure}

\newpage
\begin{figure}
\centerline{
\psfig{figure=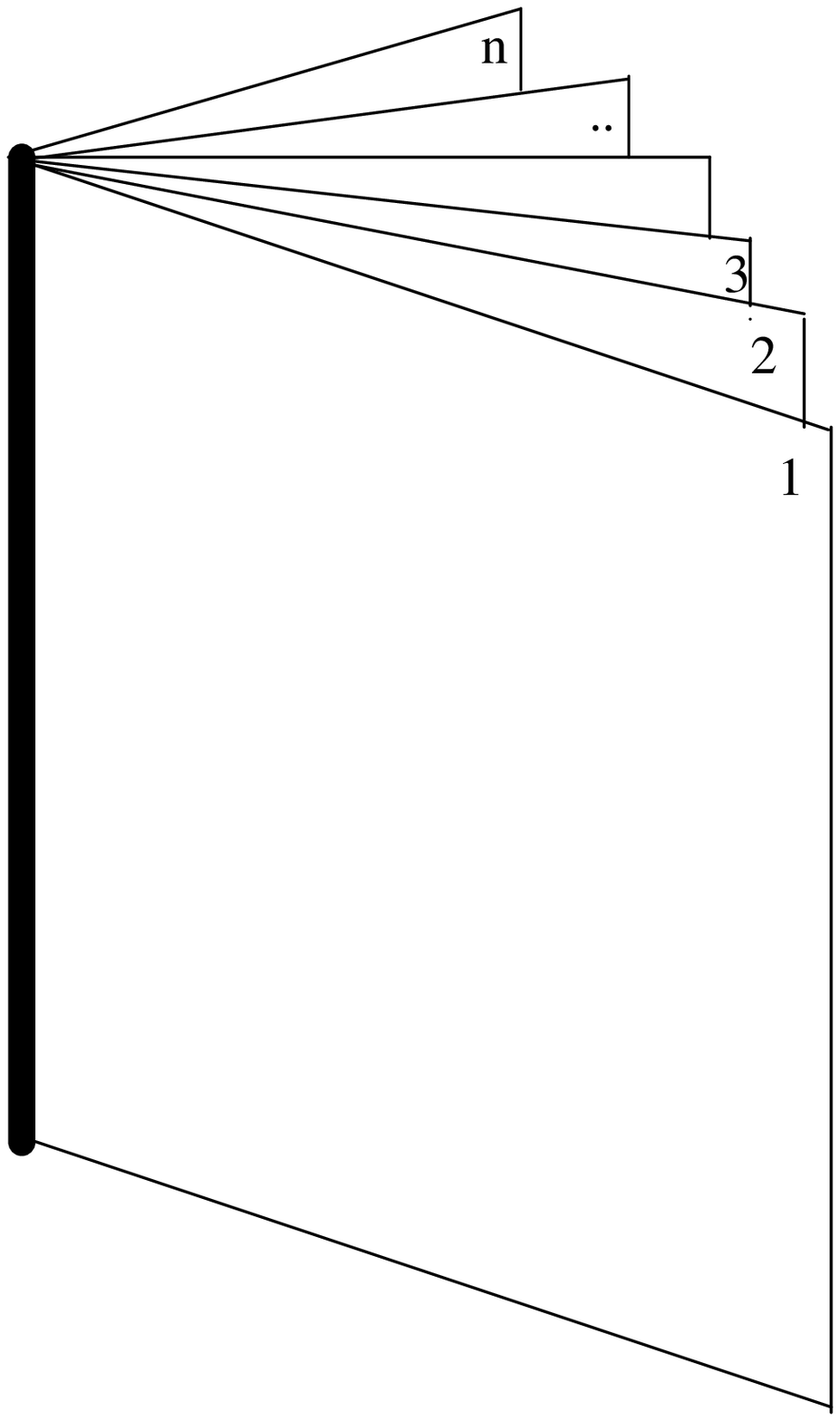}}
\vspace{3cm}
\begin{center}
{\bf Figure 2}
\end{center}
\end{figure}

\newpage
\begin{figure}
\centerline{
\psfig{figure=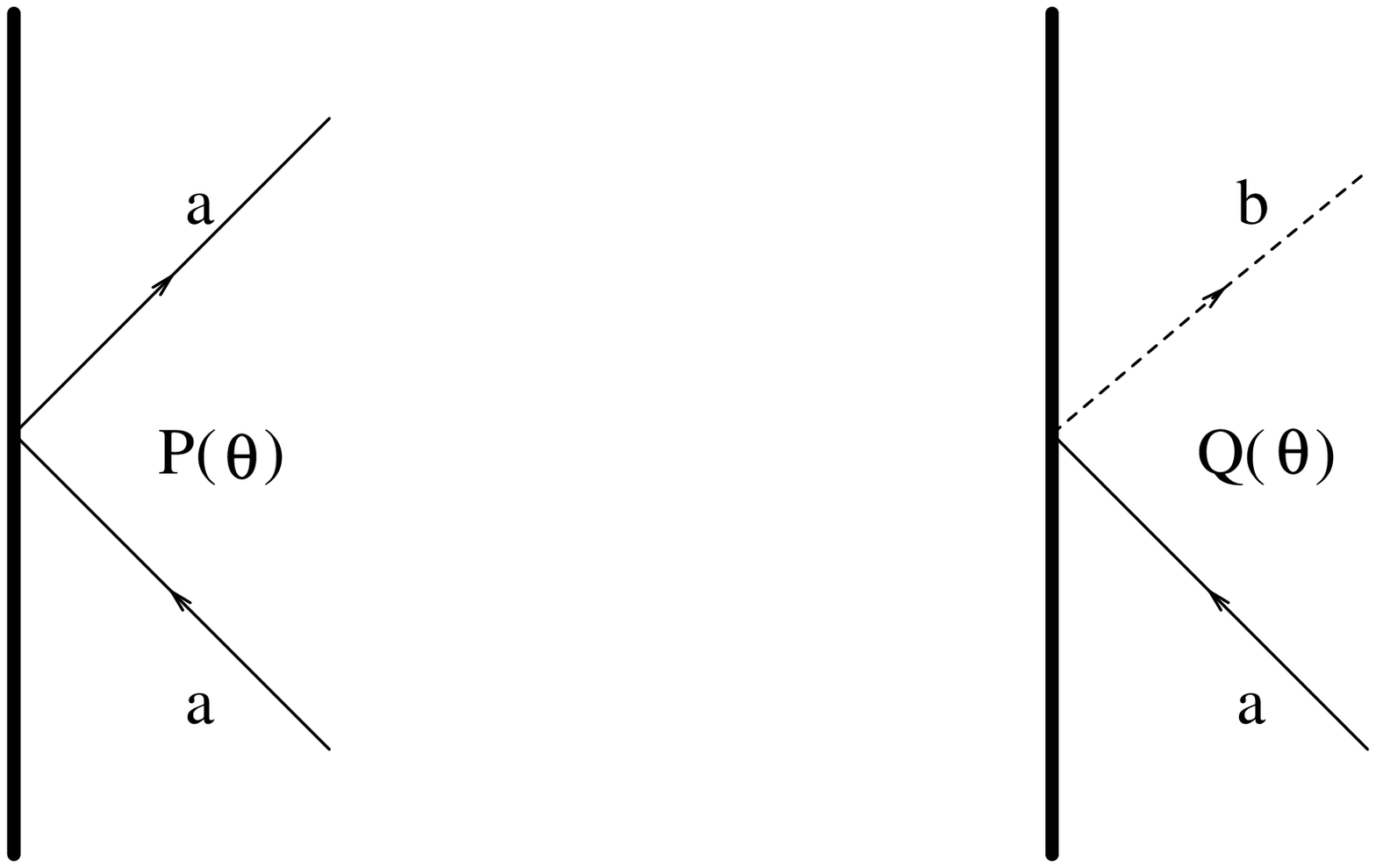}}
\vspace{3cm}
\begin{center}
{\bf Figure 3}
\end{center}
\end{figure}

\newpage
\begin{figure}
\centerline{
\psfig{figure=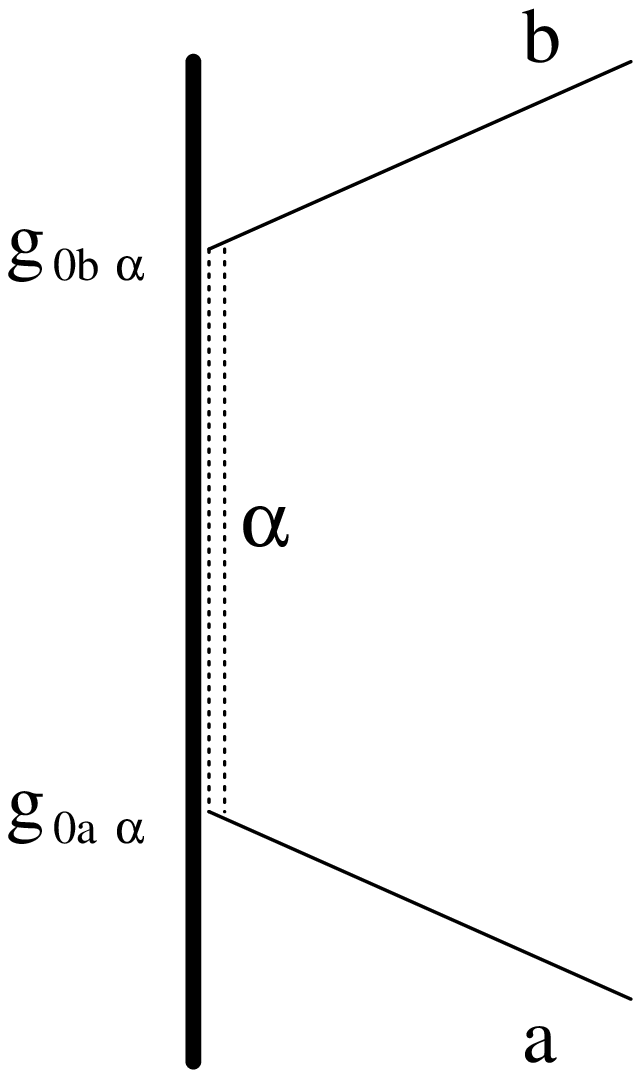}}
\vspace{3cm}
\begin{center}
{\bf Figure 4}
\end{center}
\end{figure}

\newpage
\begin{figure}
\centerline{
\psfig{figure=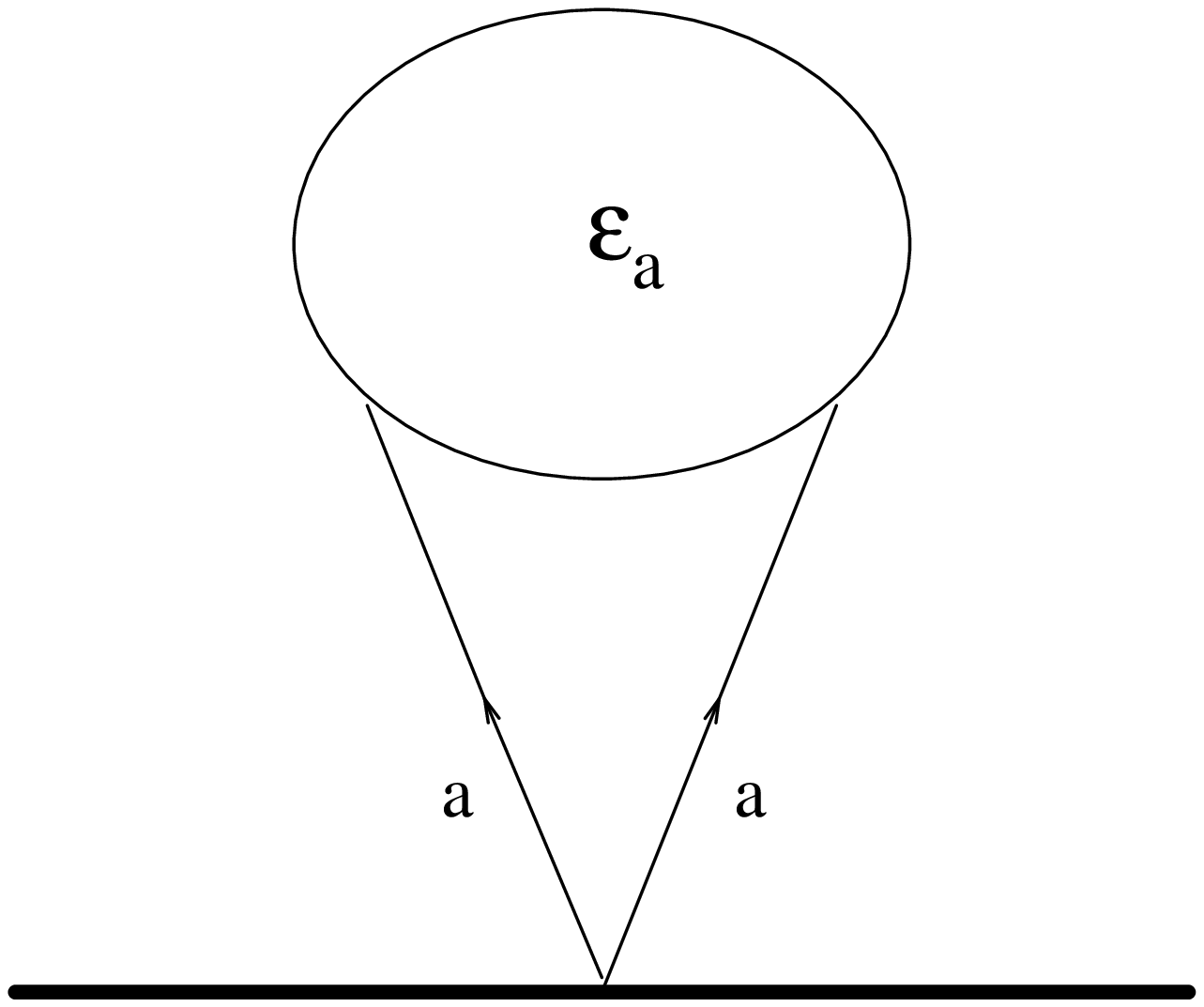}}
\vspace{3cm}
\begin{center}
{\bf Figure 5}
\end{center}
\end{figure}

\newpage
\begin{figure}
\centerline{
\psfig{figure=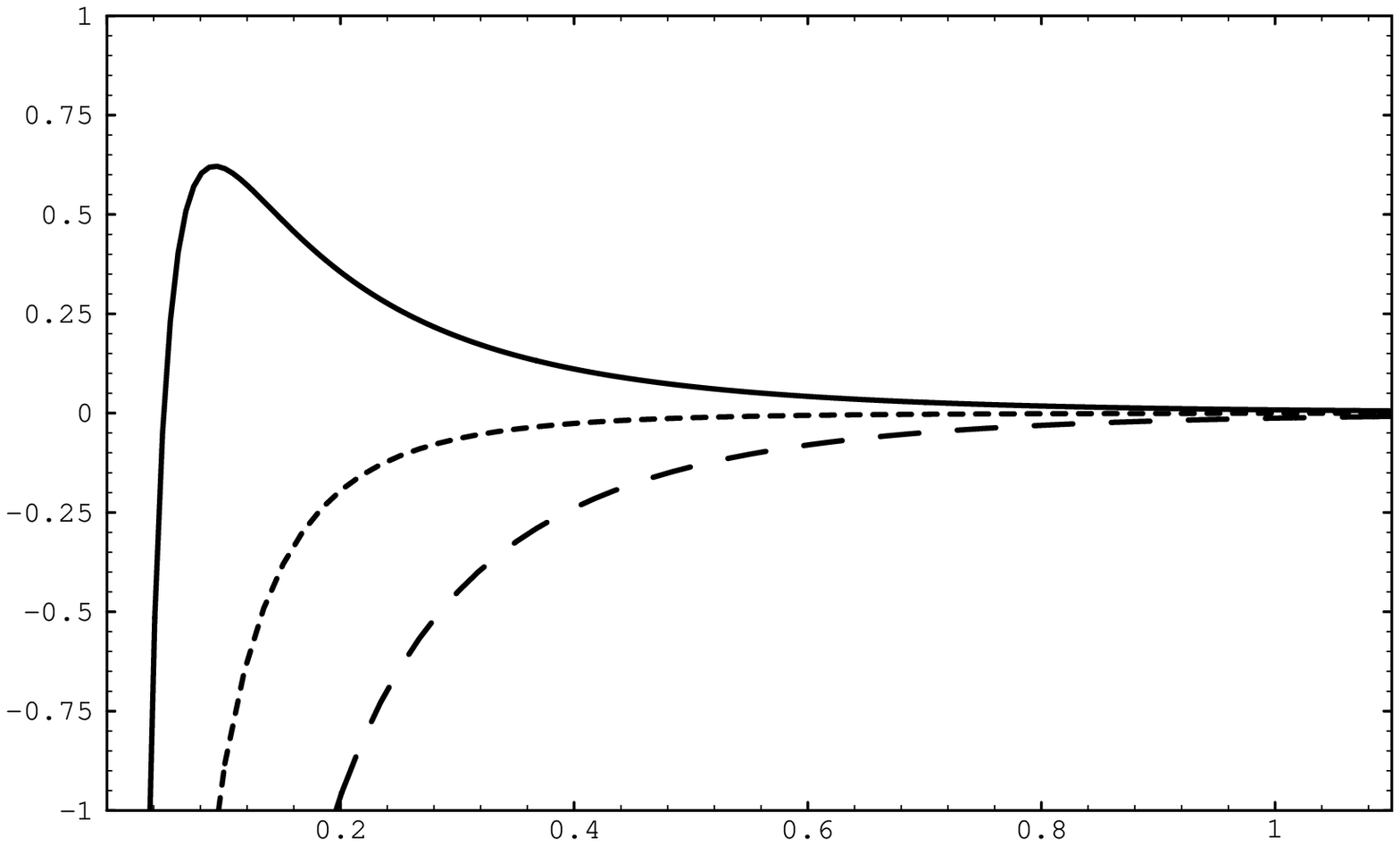}}
\begin{center}
{\bf Figure 6}
\end{center}
\end{figure}

\newpage
\begin{figure}
\centerline{
\psfig{figure=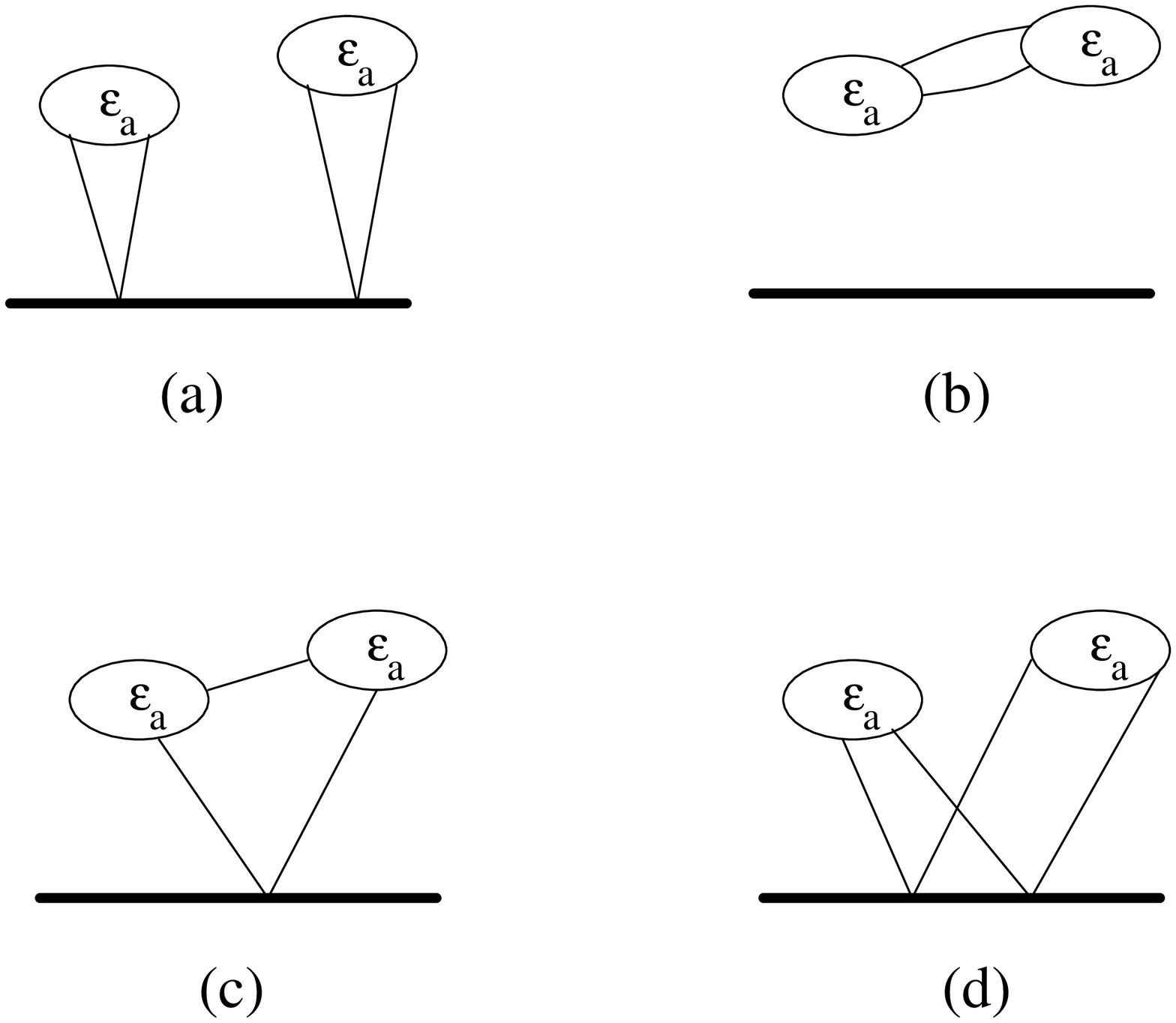}}
\vspace{3cm}
\begin{center}
{\bf Figure 7}
\end{center}
\end{figure}

\newpage
\begin{figure}
\centerline{
\psfig{figure=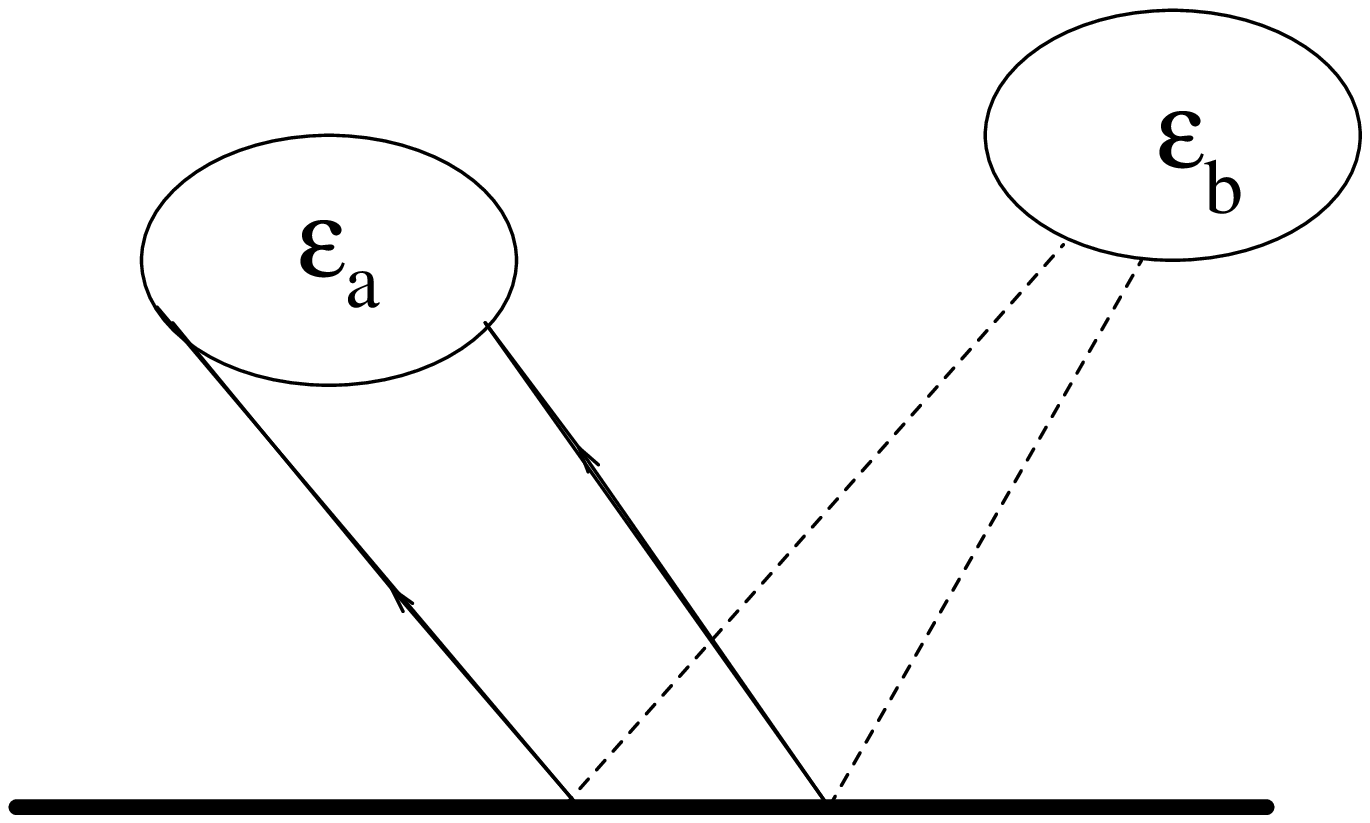}}
\vspace{3cm}
\begin{center}
{\bf Figure 8}
\end{center}
\end{figure}

\newpage
\begin{figure}
\centerline{
\psfig{figure=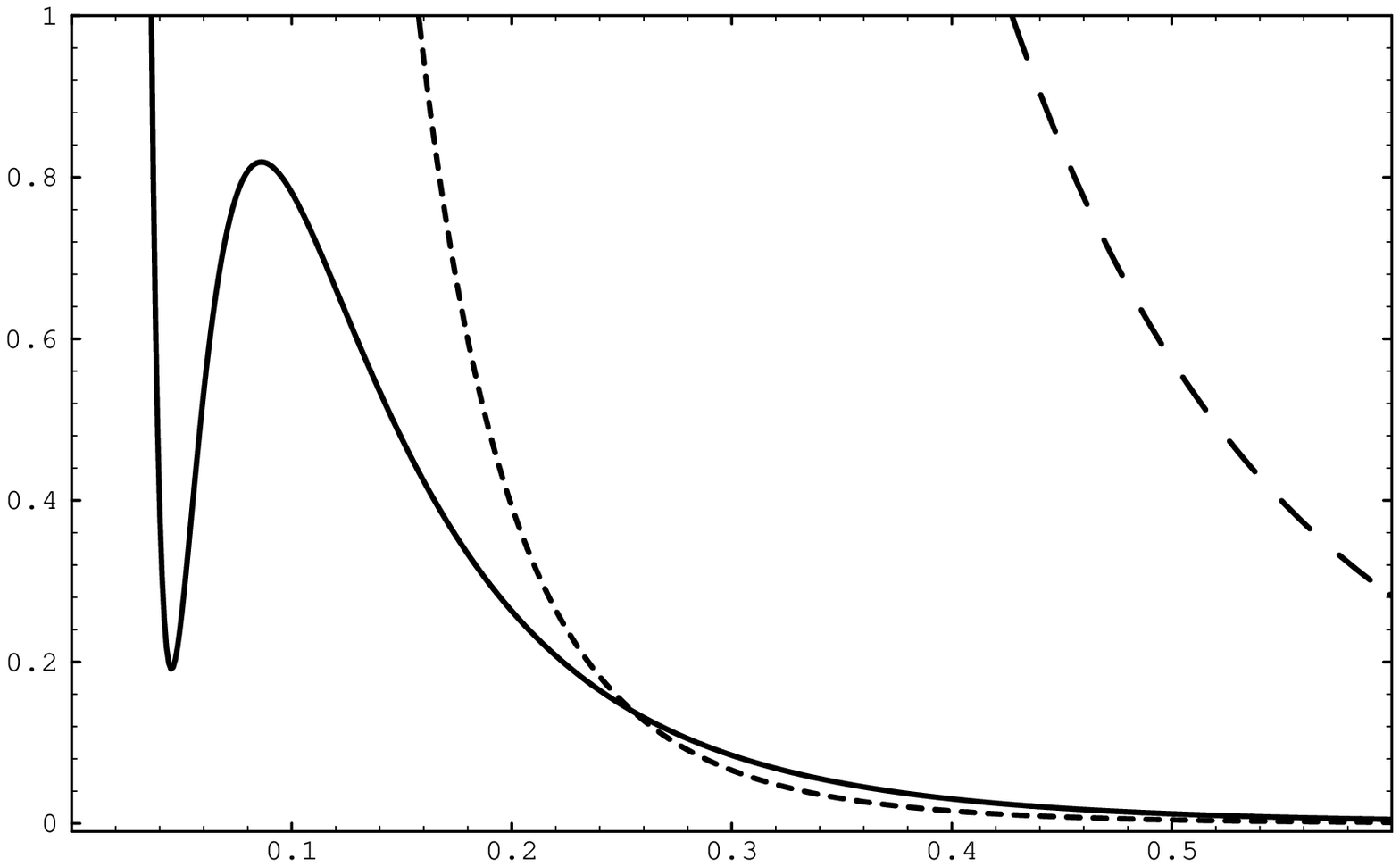}}
\begin{center}
{\bf Figure 9}
\end{center}
\end{figure}

\begin{figure}
\centerline{
\psfig{figure=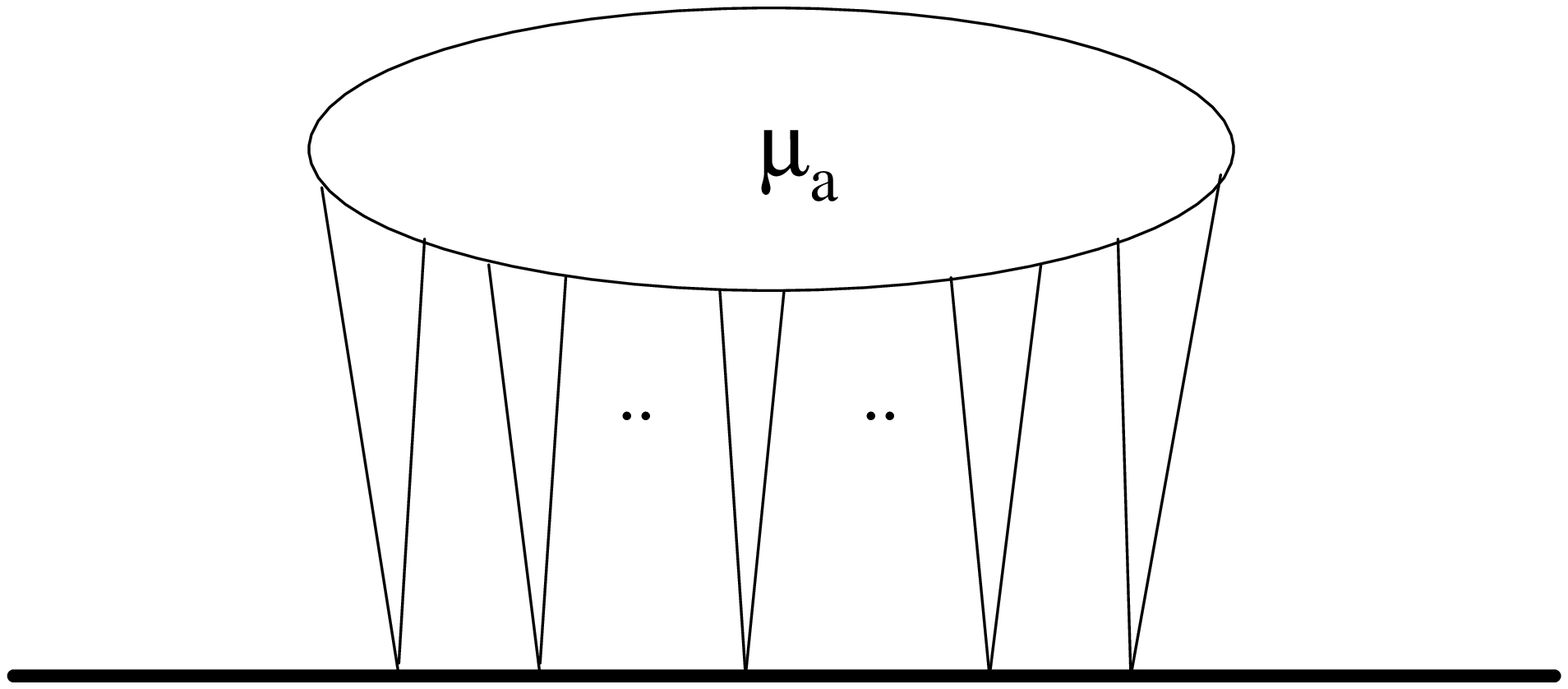}}
\vspace{3cm}  
\begin{center}
{\bf Figure 10}
\end{center}
\end{figure}

\newpage
\begin{figure}
\centerline{
\psfig{figure=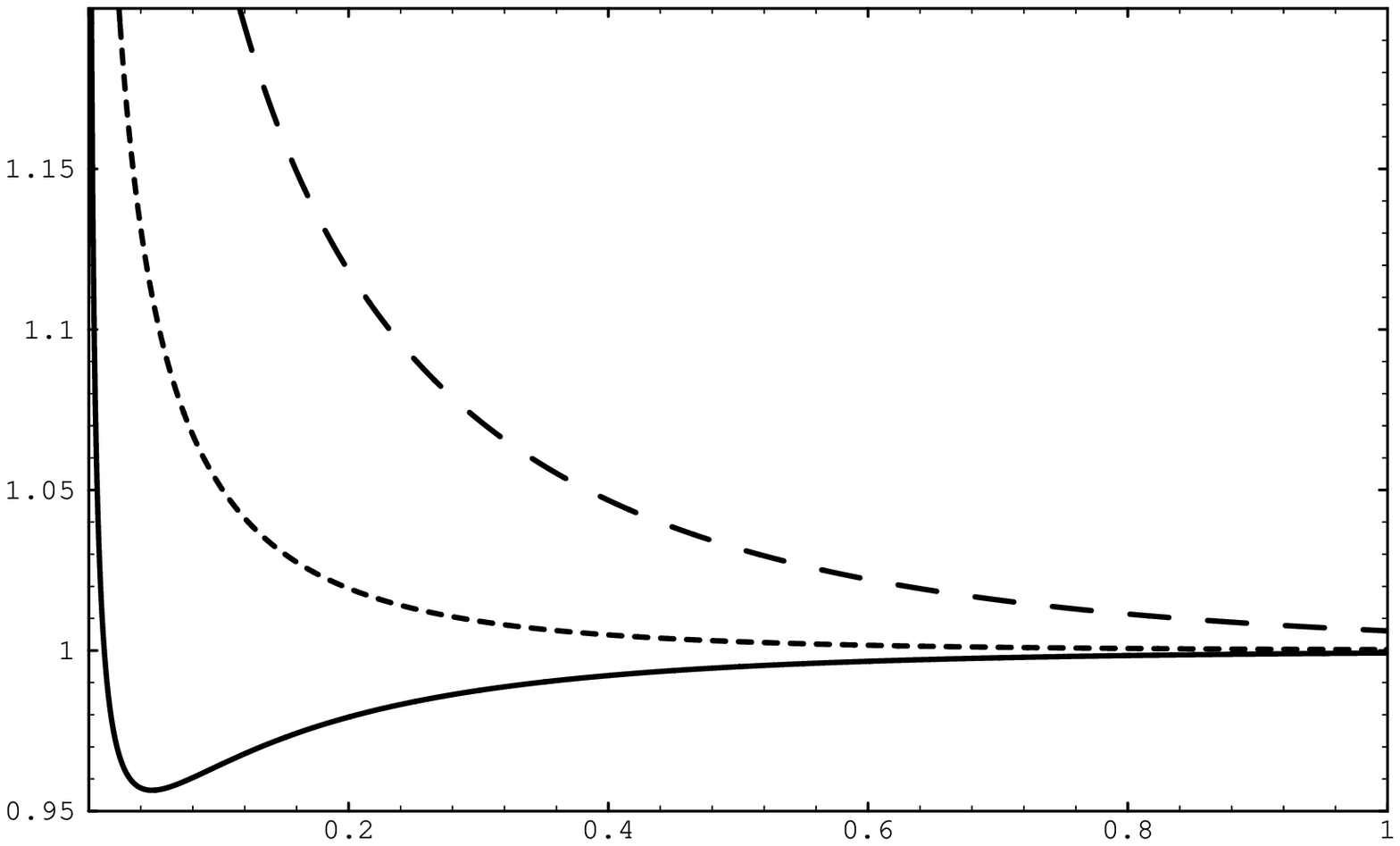}}
\begin{center}
{\bf Figure 11}
\end{center}
\end{figure}

\newpage
\begin{figure}
\centerline{
\psfig{figure=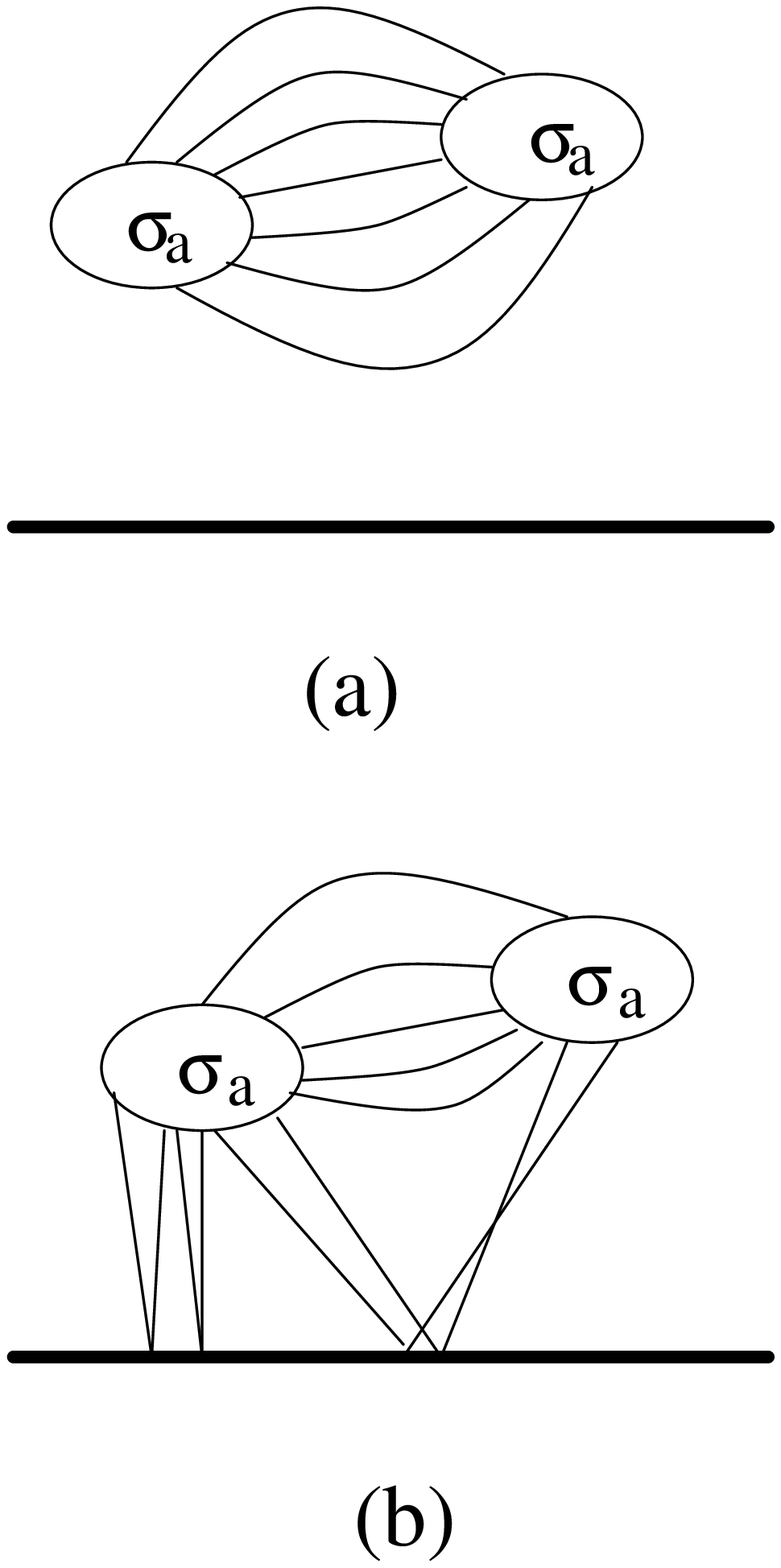}}
\vspace{3cm}
\begin{center}
{\bf Figure 12}
\end{center}
\end{figure}

\newpage
\begin{figure}
\centerline{
\psfig{figure=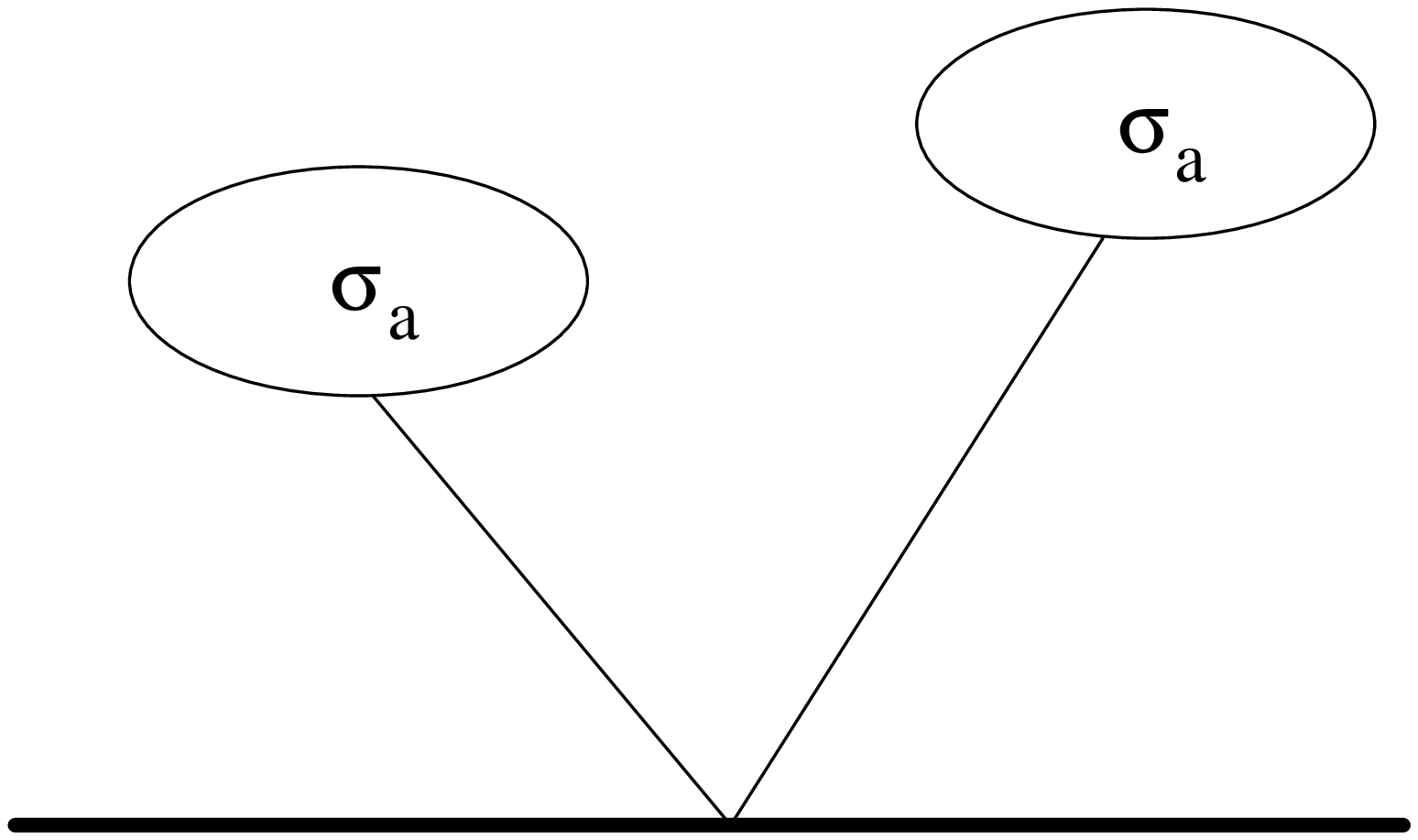}}
\vspace{3cm}
\begin{center}
{\bf Figure 13}
\end{center}
\end{figure}

\newpage
\begin{figure}
\centerline{
\psfig{figure=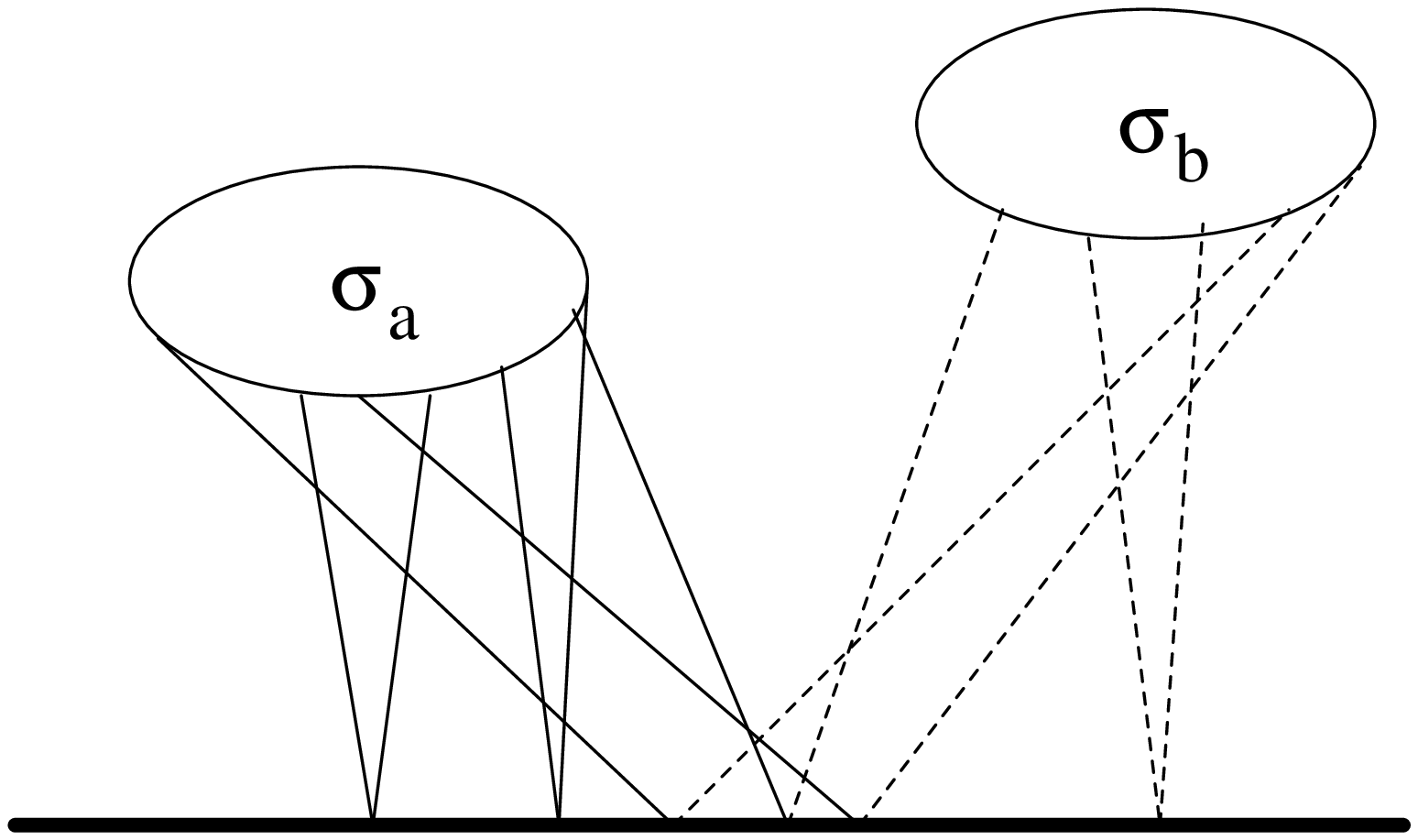}}
\vspace{3cm}
\begin{center}
{\bf Figure 14}
\end{center}
\end{figure}

\newpage
\begin{figure}
\centerline{
\psfig{figure=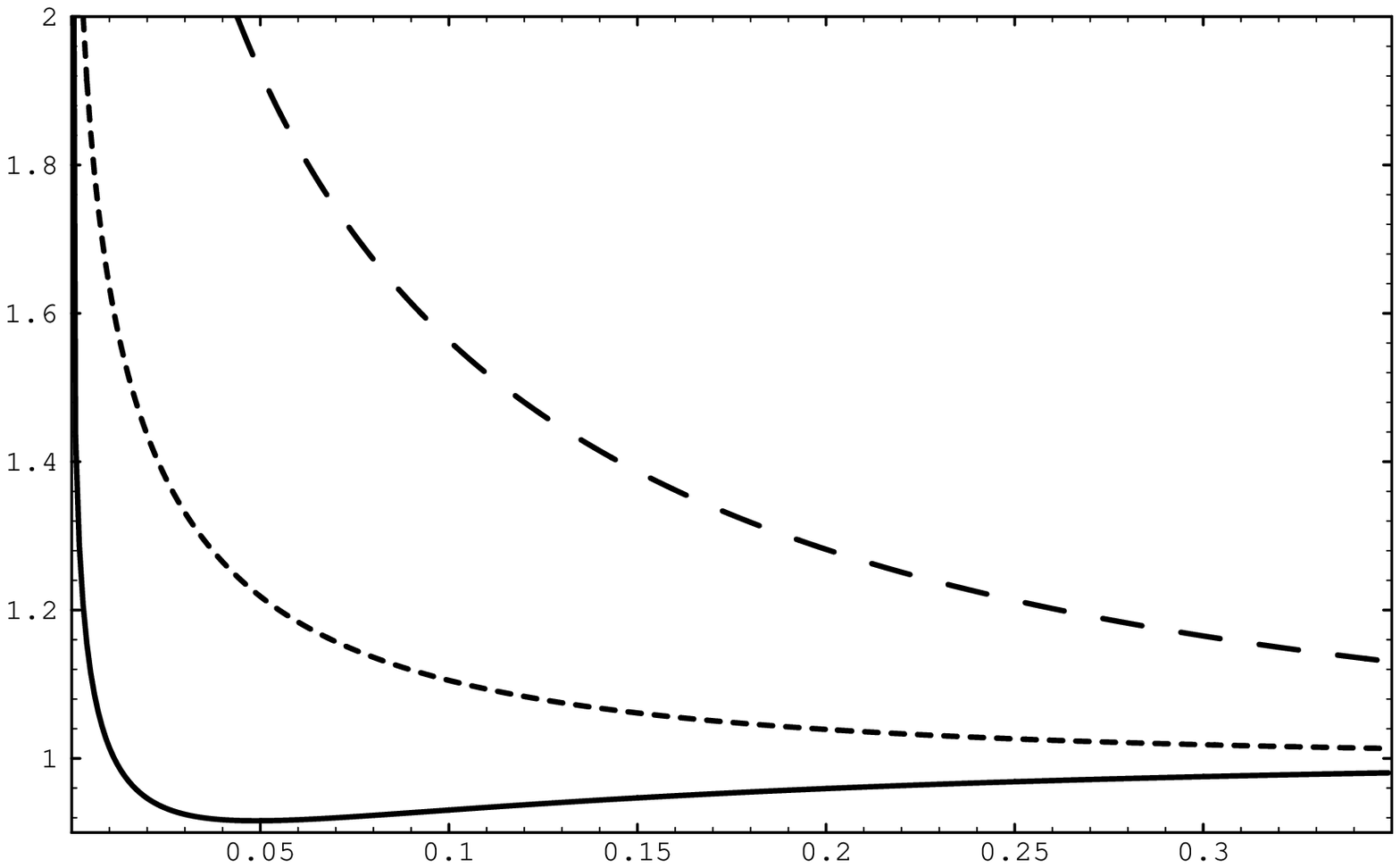}}
\begin{center}
{\bf Figure 15}
\end{center}
\end{figure}

\newpage
\begin{figure}
\centerline{
\psfig{figure=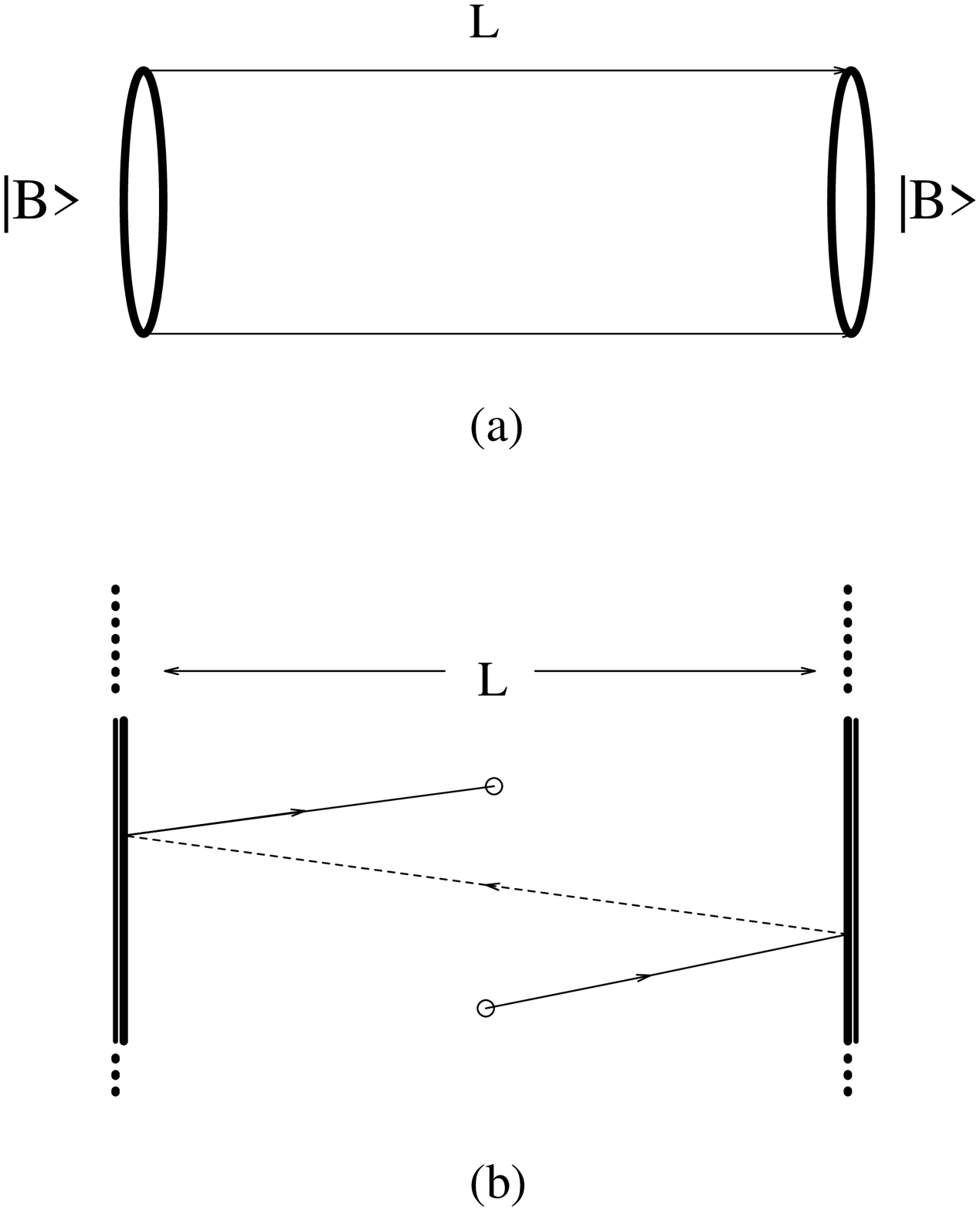}}
\begin{center}
{\bf Figure 16}
\end{center}
\end{figure}

\end{document}